\documentclass[11pt]{article}
\pdfoutput=1
\usepackage{jheppub}
\usepackage{graphicx}
\usepackage{amssymb}
\usepackage{subcaption}
\usepackage{amsmath}
\usepackage{slashed}
\usepackage{hyperref}
\usepackage{caption}
\usepackage{xcolor}
\usepackage{dsfont}

\newcommand\beq{\begin{equation}}
\newcommand\eeq{\end{equation}}

\newcommand{\Tr}{{\rm Tr\,}}

\newcommand\Dslash{D\!\!\!\!\slash}

\def\shrug{\texttt{\raisebox{0.75em}{\char`\_}\char`\\\char`\_\kern-0.5ex(\kern-0.25ex\raisebox{0.25ex}{\rotatebox{45}{\raisebox{-.75ex}"\kern-1.5ex\rotatebox{-90})}}\kern-0.5ex)\kern-0.5ex\char`\_/\raisebox{0.75em}{\char`\_}}}

\title{Novel 3d bosonic dualities from bosonization and holography}

\preprint{\today}

\author[a]{Kyle Aitken,}
\author[a]{Andrew Baumgartner,}
\author[a]{ Andreas Karch}
\affiliation[a]{Department of Physics, University of Washington, Seattle, WA, 98195-1560, USA}
\emailAdd{kaitken@uw.edu,baum4157@uw.edu,akarch@uw.edu}

\abstract{We use 3d bosonization dualities to derive new non-supersymmetric dualities between bosonic quiver theories in $2+1$ dimensions. It is shown that such dualities are a natural non-Abelian generalization of the bosonic particle-vortex duality. A special case of such dualities is applicable to Chern-Simons theories living on interfaces in $3+1$ dimensional $SU(N)$ Yang-Mills theory across which the theta angle jumps. We also analyze such interfaces in a holographic construction which provides further evidence for novel dualities between quiver gauge theories and gauge theories with adjoint scalars. These conjectured dualities pass some stringent consistency tests.
}

\begin{document}
\maketitle

\section{Overview}

Recently, a family of non-supersymmetric dualities between Chern-Simons-matter theories in $2+1$-dimensions has been conjectured \cite{Aharony:2015mjs}. Due to the fact that one side of the duality contains bosons and the other has fermions, such identifications have been termed ``3d bosonization'' or ``Aharony's dualities''. Schematically, these dualities state
\begin{subequations}
\label{eq:aharony_schematic}
\begin{align}
SU(k)_N\,\text{with \ensuremath{N_f} \ensuremath{\phi}}\qquad & \leftrightarrow\qquad U(N)_{-k+\frac{N_f}{2}}\,\text{with \ensuremath{N_{f}} \ensuremath{\psi}},\label{eq:u ferm tw}\\
U(k)_{N}\,\text{with \ensuremath{N_{f}} \ensuremath{\phi}}\qquad & \leftrightarrow\qquad SU(N)_{-k+\frac{N_{f}}{2}}\,\text{with \ensuremath{N_{f}} \ensuremath{\psi}}\label{eq:u scalar tw}
\end{align}
\end{subequations}
where $\phi$ are self-interacting scalars, $\psi$ are free fermions, and  ``$\leftrightarrow$'' means the theories share an IR fixed point. These dualities are subject to the flavor bound $N_f\leq k$.

The strongest evidence for such dualities is based on studies where the level ($k$) and the rank ($N$) are taken to be much greater than one (but $k/N$ is held fixed). In this limit observables are under perturbative control \cite{Giombi:2011kc, Aharony:2011jz, Aharony:2012nh, Jain:2014nza, Inbasekar:2015tsa, Minwalla:2015sca, Gur-Ari:2016xff} and one can confirm that many observables on both sides of the duality such as the operator spectrum, free energy, and correlation functions match to leading order. Additionally, one can deform away from the IR fixed point by including relevant operators in the Lagrangian. This procedure yields topological field theories (TFTs) which are level-rank dual and hence equivalent.

Surprisingly, further evidence for these dualities arises in the exact opposite regime, where $N=k=N_f=1$. In this case, \eqref{eq:aharony_schematic} reduces to
\begin{subequations}
\label{eq:abelian_aharony}
\begin{align}
\text{Wilson-Fisher scalar} \qquad & \leftrightarrow\qquad U(1)_{-1/2}+\text{fermion},\\
U(1)_{1}+\text{scalar}\qquad & \leftrightarrow\qquad \text{free fermion}.
\end{align}
\end{subequations}
These ``Abelian dualities'' have been used to derive an entire web of related dualities \cite{Karch:2016sxi, Seiberg:2016gmd}, within which is the well-known bosonic particle-vortex duality \cite{Peskin:1977kp, Dasgupta:1981zz} and its recently discovered fermionic equivalent \cite{Son:2015xqa}. The methodology used in deriving this web of Abelian dualities has been extended to Abelian and non-Abelian linear quivers \cite{Karch:2016aux, Jensen:2017dso} to generate even more novel dualities, although these are often limited in scope due to flavor bounds.

More recently, a generalization of Aharony's dualities has been discovered \cite{Benini:2017aed,Jensen:2017bjo} where each side of the duality has fermions and scalars,
\begin{equation}
SU(N)_{-k+\frac{N_{f}}{2}}\,\text{with \ensuremath{N_{s}} \ensuremath{\phi}\ and \ensuremath{N_{f}} \ensuremath{\psi}}\qquad\leftrightarrow\qquad U(k)_{N-\frac{N_{s}}{2}}\,\text{with \ensuremath{N_{f}} \ensuremath{\Phi}\ and \ensuremath{N_{s}} \ensuremath{\Psi}}.\label{eq:master tw}
\end{equation}
Note that this duality reduces to \eqref{eq:u ferm tw} and \eqref{eq:u scalar tw} when $N_f=0$ and $N_s=0$, respectively. We will refer to this duality as the ``master duality'' since \eqref{eq:aharony_schematic} can be recovered as a special case. Novel to the master duality is the fact the scalar and fermionic matter on each side of the duality interact with one another through a quartic term and each type of matter is subject to its own flavor bound.

Said dualities have application toward the half-filled fractional quantum Hall effect as well as  surface states of topological insulators \cite{Son:2015xqa, wang2015dual, Metlitski:2015eka, Seiberg:2016gmd}. Further support for these dualities include deformations from supersymmetric cases \cite{Jain:2013gza, Gur-Ari:2015pca, Kachru:2016rui, Kachru:2016aon}, derivation from an array of $1+1$ dimensional wires \cite{mross2016explicit}, a matching of global symmetries and 't Hooft anomalies \cite{Benini:2017dus}, consistency checks of the dualities on manifolds with boundaries \cite{Aitken:2017nfd, Aitken:2018joi}, and support from Euclidean lattice constructions \cite{Chen:2017lkr, Chen:2018vmz, Jian:2018amu}.

In this work we use the master duality and methods similar to those developed in \cite{Jensen:2017dso} to derive novel Bose-Bose dualities between non-Abelian linear quivers. We argue that these dualities can be viewed as a natural generalization of the bosonic particle-vortex duality to non-Abelian gauge groups since the quivers share many of the qualitative features present in the particle-vortex duality.

Of particular interest is the application of these dualities to $2+1$-dimensional defects in Yang-Mills theory on $\mathbb{R}^4$, which will be the focus of the latter half of this paper. It has recently been shown that there is a mixed 't Hooft anomaly between time-reversal symmetry and center symmetry at $\theta=\pi$ \cite{Gaiotto:2017yup}.
This is rooted in the fact that $SU(N)$ YM theory is believed to have $N$ distinct vacua associated to $N$ branches of the theory. Such branches are individually $2\pi N$ periodic and correspond to $SU(N)/\mathbb{Z}_N$ gauge theories. This seems to contradict the long held belief that $\theta$ is $2\pi$ periodic in $SU(N)$ YM theory, but the conflict is resolved since the vacua interchange roles under a $2\pi$ transformation. More specifically, if one tracks the true ground state of the theory, one changes branches in a single $2\pi$ period. Thus, as theta is varied from, say $\theta=0$ to $2\pi n$, the theory traverses several vacua. However, this changes when one couples the one-form center symmetry to a background (two-form) gauge field. In this case one cannot consistently choose the coefficient of the counterterm, sometimes referred to as the ``discrete theta angle'', to make the theory non-anomalous. Since this counterterm changes as one traverses branches, a spatially varying $\theta$ angle gives rise to domain walls separating regions with distinct discrete theta angle.  Using anomaly inflow arguments, the effective field theory living on the interface is found to be a Chern-Simons gauge theory (see \cite{Gaiotto:2017yup, Gaiotto:2017tne} for more details).

Although anomaly considerations require a non-trivial theory to live on the interface, they alone do not fully fix the theory. Among others, $[SU(N)_{-1}]^n$ or $SU(N)_{-n}$ would be consistent choices.\footnote{Note that we are changing the direction of the $\theta$ gradient relative to \cite{Gaiotto:2017tne} and so have negative levels for our  Chern-Simons theories. This is in order to conform to the conventions of \cite{Jensen:2017xbs} for the stringy embeddings.} The authors of \cite{Gaiotto:2017tne} argue that, at least at $n \ll N$,  $[SU(N)_{-1}]^n$ is the appropriate description for slowly varying theta (meaning that $|\nabla \theta| \ll \Lambda$ where $\Lambda$ is the strong coupling scale of the confining gauge theory), whereas $SU(N)_{-n}$ is appropriate for a sharp interface such as a discrete jump by $2 \pi n$ at a given location. If these are indeed the correct descriptions this suggests that there is a phase transition as one smooths out a given jump in $\theta$. If this phase transition is second order, the transition point would be governed by a CFT which is most easily realized as a Chern-Simons-matter theory. In any case, this CFT can serve as a parent theory from which topological field theories, describing either the slowly varying as well as the sharp step, can be realized as massive deformation.

\begin{figure}
\centering
\includegraphics[scale=0.7]{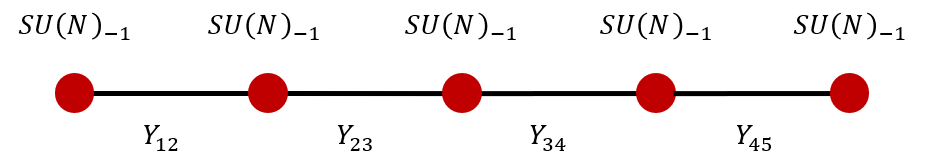}
\caption{Parent  Chern-Simons  matter theory at the phase transition for the special case $n=5$. Nodes represent gauge theories with the associated Chern-Simons term and links represent matter bifundamentally charged under the gauge groups on the adjascent nodes.} \label{fig:parentcft}
\end{figure}

The conjectured CFT between the two extreme phases is schematically
\begin{align}
\label{eq:su_quiver_ym}
[SU(N)_{-1}]^n + \text{bifundamental scalars}
\end{align}
which was used in ref. \cite{Gaiotto:2017tne} to explain the transition between two different vacua of $\left(3+1\right)$-dimensional Yang-Mills. This parent CFT is based on a quiver gauge theory as displayed in Fig. \ref{fig:parentcft}. Each node depicts a $SU(N)_{-1}$  Chern-Simons  gauge theory, the links connecting them represent bifundamental scalar fields, $Y$. The theory has two obvious massive deformations: we can give all the scalars a positive or a negative mass squared. In the former case the scalars simply decouple and we are left with the $[SU(N)_{-1}]^n$ TFT appropriate for slowly varying theta, in the latter case the gauge group factors get Higgsed down to the diagonal subgroup and we find the $SU(N)_{-n}$ associated with the steep defect. There are also mixed phases, where some of the $Y$ have negative and some positive mass squared.

In this work, we propose a theory dual to \eqref{eq:su_quiver_ym} which is supported by both 3d bosonization of non-Abelian linear quivers and holographic duality. The proposed ``theta wall'' duality is
\begin{align}
\label{eq:u_quiver_ym}
[SU(N)_{-1}]^n + \text{bifundamental scalars} \qquad\leftrightarrow\qquad U(n)_{N} + \text{adjoint scalars}.
\end{align}
We will see that this is a special case of the more general quiver dualities derived in Sec. \ref{sec:quivers} which do not include matter in the adjoint. This is a special feature of \eqref{eq:u_quiver_ym}, owed to the fact that when all ranks of the $SU$ quiver theory are equal, the $U$ quiver contains nodes which are confining. With the careful addition of interactions in the proposed theories, mass deformations on either side of the duality yield TFTs which are level-rank dual to each other.

The paper is outlined as follows. In Sec. \ref{sec:review} we review the master duality and establish the conventions we use for the rest of the paper. Sec. \ref{sec:quivers} contains our derivation of the non-Abelian linear quiver dualities, including the details of how such dualities should be viewed as generalization of the particle-vortex duality. We then specialize to quivers applicable to theta interfaces in $3+1$-dimensional $SU(N)$ Yang-Mills theory in Sec. \ref{sec:thetawalls}. Subsections \ref{sec:thetawall_3d} and \ref{sec:thetawall_holo} contain the 3d bosonization and holographic support for such dualities, respectively. In Sec. \ref{sec:conclusion} we discuss our results and conclude. The appendix contains several details of our construction of the non-Abelian quivers.

As we were finalizing this work, we were made aware of \cite{Argurio:2018uup} which studies domain walls in different phases of the Witten-Sakai-Sugimoto model. This has some overlap with Sec. \ref{sec:jensen}, particularly regarding the nature of domain walls in the pure YM sector.

\section{Review of 3d Bosonization}
\label{sec:review}

We begin by reviewing 3d bosonization and establish conventions we will use throughout this paper. The most general form of 3d bosonization, the so-called master bosonization duality \cite{Benini:2017aed,Jensen:2017bjo}, is a conjecture that the following two Lagrangians share the same IR fixed point\footnote{Here we follow the conventions outlined in ref. \cite{Aitken:2018joi}. We have dropped all gravitational Chern-Simons terms since they are not relevant for our purposes. Note there is a slight difference in convention in the sign of the BF term and the $\tilde{A}_2$ coupling on the $U$ side of the duality. However since the difference always amounts to an even number of sign changes the TFTs still match under mass deformations. Additionally, the flux attachment procedure picks up two minus signs from this effect as well, meaning the quantum numbers of the baryon and monopole operators still match.}
\begin{subequations}
\label{eq:master_dual}
\begin{align}
\mathcal{L}_{SU} & =\left|D_{b^{\prime}+B+\tilde{A}_{1}+\tilde{A}_{2}}\phi\right|^{2}+i\bar{\psi}\Dslash_{b^{\prime}+C+\tilde{A}_{1}}\psi+\mathcal{L}_{\text{int}}-i\left[\frac{N_{f}-k}{4\pi}\text{Tr}_{N}\left(b^{\prime}db^{\prime}-i\frac{2}{3}b^{\prime3}\right)\right]\nonumber \\
 & -i\left[\frac{N}{4\pi}\text{Tr}_{N_{f}}\left(CdC-i\frac{2}{3}C^{3}\right)+\frac{N(N_{f}-k)}{4\pi}\tilde{A}_{1}d\tilde{A}_{1}\right],\\
\mathcal{L}_{U} & =\left|D_{c+C}\Phi\right|^{2}+i\bar{\Psi}\Dslash_{c+B+\tilde{A}_{2}}\Psi+\mathcal{L}_{\text{int}}^{\prime}-i\left[\frac{N}{4\pi}\text{Tr}_{k}\left(cdc-i\frac{2}{3}c^{3}\right)-\frac{N}{2\pi}\text{Tr}_{k}(c)d\tilde{A}_{1}\right] \label{eq:master_LU}
\end{align}
\end{subequations}
with the mass identifications $m_\psi\leftrightarrow -m_\Phi^2$ and $m_\phi^2\leftrightarrow m_\Psi$. Our definitions of fields are shown in Table \ref{tab:master_notation}. We will use uppercase letters for background gauge fields, lowercase for dynamical gauge fields, and Abelian fields carry a tilde. This duality is subject to the flavor bound $(N_f,N_s)\leq(k,N)$, but excludes the case $(N_f,N_s)=(k,N)$.\footnote{There are proposals for dualities describing the phase structure of these theories slightly beyond the bounds \cite{Komargodski:2017keh}, but such cases will not be relevant for this work.} Our notation for covariant derivatives is
\begin{subequations}
\begin{align}
\left(D_{b^{\prime}+B+\tilde{A}_{1}+\tilde{A}_{2}}\right)_{\mu}\phi & =\left[\partial_{\mu}-i\left(b_{\mu}^{\prime}\mathds{1}_{N_s}+B_{\mu}\mathds{1}_{N}+\tilde{A}_{1\mu}\mathds{1}_{N N_s}+\tilde{A}_{2\mu}\mathds{1}_{N N_s}\right)\right]\phi,\\
\left(D_{b^{\prime}+C+\tilde{A}_1}\right)_{\mu}\psi & =\left[\partial_{\mu}-i\left(b_{\mu}^{\prime}\mathds{1}_{N_f}+C_{\mu}\mathds{1}_{N}+\tilde{A}_{1\mu}\mathds{1}_{N N_f}\right)\right]\psi,\\
\left(D_{c+C}\right)_{\mu}\Phi & =\left[\partial_{\mu}-i\left(c_{\mu}\mathds{1}_{N_f}+C_{\mu}\mathds{1}_{k}\right)\right]\Phi,\\
\left(D_{c+B+\tilde{A}_{2}}\right)_{\mu}\Psi & =\left[\partial_{\mu}-i\left(c_{\mu}\mathds{1}_{N_s}+B_{\mu}\mathds{1}_{k}+\tilde{A}_{2\mu}\mathds{1}_{k N_s}\right)\right]\Psi.
\end{align}
\end{subequations}
The interaction terms are
\begin{subequations}
\label{eq:master_ints}
\begin{align}
\mathcal{L}_{\text{int}} & =\alpha\left(\phi^{\dagger a_{c}a_{s}}\phi_{a_{c}a_{s}}\right)^{2}-C \left(\bar{\psi}^{a_{c}a_{f}}\phi_{a_{c}a_{s}}\right)\left(\phi^{\dagger b_{c}a_{s}}\psi_{b_{c}a_{f}}\right)\label{eq: master ints 1 tw}\\
\mathcal{L}_{\text{int}}^{\prime} & =\alpha\left(\Phi^{\dagger a_{c}a_{f}}\Phi_{a_{c}a_{f}}\right)^{2}+C' \left(\bar{\Psi}^{a_{c}a_{s}}\Phi_{a_{c}a_{f}}\right)\left(\Phi^{\dagger b_{c}a_{f}}\Psi_{b_{c}a_{s}}\right)\label{eq: master ints 2 tw}
\end{align}
\end{subequations}
where $a_c,b_c$ are indices associated with the color symmetries; $a_f,b_f$ with the $SU(N_f)$ symmetry; and $a_s,b_s$ with the $SU(N_s)$ symmetry. $C$ and $C^\prime$ coefficients of the associated interactions which we will later fix. The quartic scalar terms will henceforth be implied anytime a scalar is present, but we will make note of the scalar/fermion interaction terms when they exist.

\begin{table}
\begin{centering}
\begin{tabular}{|c|c|c||c|c|c|c|}
\cline{2-7}
\multicolumn{1}{c|}{}& \multicolumn{2}{c||}{Gauge Fields} & \multicolumn{4}{c |}{Background Fields} \\
\hline
\textbf{Symmetry} &$SU(N)$& $U(k)$ &$SU(N_s)$ & $SU(N_f)$& $U(1)_{m,b}$& $U(1)_{F,S}$
\tabularnewline
\hline
\textbf{Field} & $b^\prime_\mu$ & $c_\mu$ & $B_\mu$ & $C_\mu$& $\tilde{A}_{1\mu}$&  $\tilde{A}_{2\mu}$
\tabularnewline
\hline
\end{tabular}
\par\end{centering}
\caption{Various gauge fields used in the master duality. Dynamical fields are denoted by lowercase letters while background fields by uppercase. $\tilde{A}_{1\mu}$ is associated with the monopole/baryon number $U(1)$ symmetry also present in Aharony's dualities. $\tilde{A}_{2\mu}$ is associated to the $U(1)$ symmetry which couples to the additional fermion/scalar matter in the master duality. \label{tab:master_notation}}
\end{table}

As mentioned in the introduction, Aharony's dualities \eqref{eq:aharony_schematic} can be found by taking the $N_s=0$ and $N_f=0$ limits of \eqref{eq:master_dual}. For example, Aharony's duality \eqref{eq:u ferm tw} is the $N_f=0$ limit and is an IR duality between Lagrangians
\begin{subequations}
\label{eq:aharony_dual}
\begin{align}
\mathcal{L}_{SU} & = \left| D_{b^\prime+B+\tilde{A}_1}\phi \right|^2 -i\left[-\frac{k}{4\pi}\text{Tr}_{N}\left(b^{\prime}db^{\prime}-i\frac{2}{3}b^{\prime 3}\right)-\frac{N k}{4\pi}\tilde{A}_{1}d\tilde{A}_{1}\right],\label{eq:mdb lsu-1-1}\\
\mathcal{L}_{U} & = i\bar{\Psi}\Dslash_{c+B}\Psi-i\left[\frac{N}{4\pi}\text{Tr}_{k}\left(cdc-i\frac{2}{3}c^{3}\right)-\frac{N}{2\pi}\text{Tr}_{k}(c)d\tilde{A}_1 \right]\label{eq:mdb lu-1-1}
\end{align}
\end{subequations}
which are subject to the flavor bounds $N_s\leq N$.

We will use the $\eta$-invariant convention where a positive mass deformation for the fermion will not change the level of the Chern-Simons term. When compared to the often employed convention where an ill defined naive Dirac operator gets augmented with an half integer Chern-Simons term this means we replace \cite{Witten:2015aba,Witten:2015aoa}:
\begin{align}
i\bar{\psi}\Dslash_A \psi - i \left[- \frac{N_f}{8\pi} \text{Tr}_N  \left( AdA-i\frac{2}{3}A^3 \right)\right] \qquad \to \qquad i\bar{\psi}\Dslash_A \psi.
\end{align}
We will continue to denote fermion half-levels when specifying the Chern-Simons theory.

\section{Non-Abelian Linear Quiver Dualities}
\label{sec:quivers}

We now turn to constructing linear quivers using the master duality. As explained in the introduction, we are ultimately motivated by the theta wall construction that leads to \eqref{eq:u_quiver_ym}, but we will derive dualities for a far more general case. We will begin with recasting the 3d bosonization derivation of  bosonic particle-vortex duality in a way that highlights the relation to the non-Abelian quivers.

\subsection{Bosonic Particle-Vortex Duality}
\label{sec:boson_pv}
\begin{figure}
\begin{centering}
\includegraphics[scale=0.6]{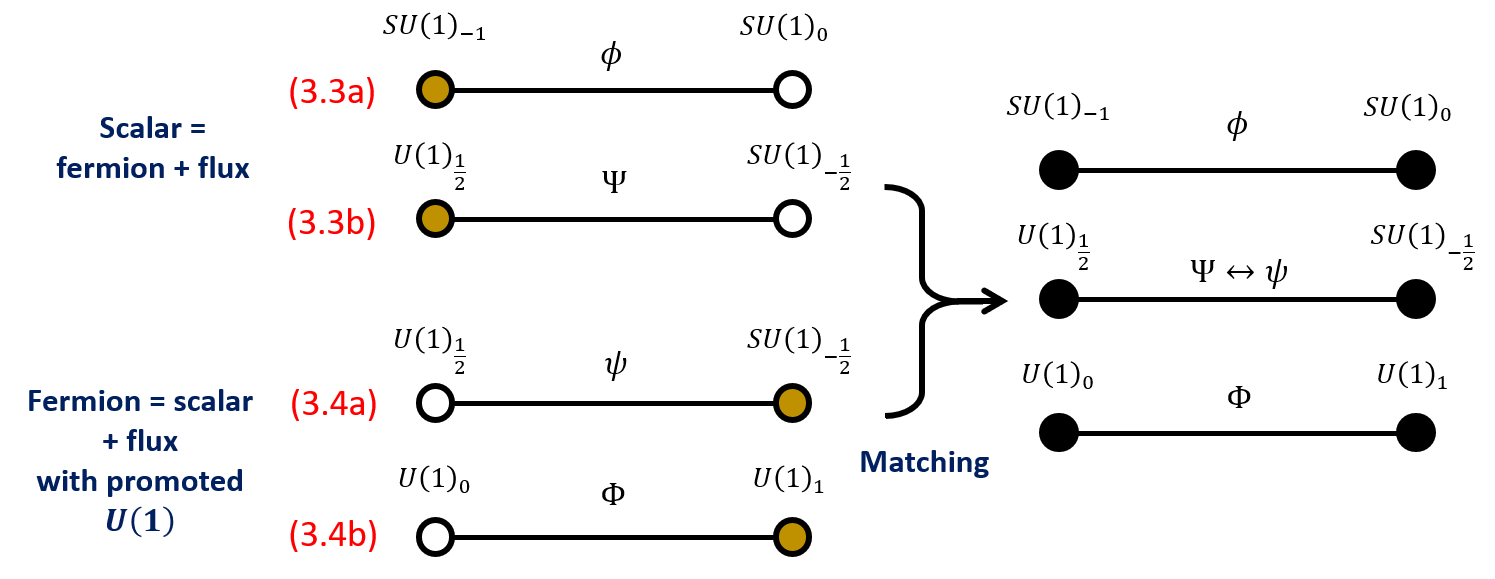}
\par\end{centering}
\caption{Derivation of the bosonic particle-vortex duality as a duality between two-node linear quiver theories. On the left-hand side, we have represented each side of Aharony's Abelian dualities as a two-node quiver. The filled yellow circle represents the color gauge group while the empty circle represents promoted global symmetries (which for the case of $SU(1)$ are placeholders). The equation numbers corresponding to the two-node quivers are shown in red.  Since the two fermionic theories are the same, one can perform a matching to arrive at a duality between three two-node quiver theories, the top and bottom of which are the XY and Abelian Higgs models, respectively. \label{fig:pv_dual}}
\end{figure}

To derive the bosonic particle-vortex duality we will use 3d bosonization techniques similar to those used in refs. \cite{Karch:2016sxi,Seiberg:2016gmd}.  We then show how one can reinterpret the derivation in terms of a two-node quiver. This will be the simplest non-trivial case of the far more general quivers we derive in Sec. \ref{sec:building_quivers}. We will drop tildes from Abelian gauge fields in this subsection since the distinction is not necessary.

Recall the bosonic particle-vortex duality states that, at low energies, the XY model is dual to the Abelian Higgs model \cite{Peskin:1977kp, Dasgupta:1981zz},
\begin{align}
\label{eq:pv_dual}
\mathcal{L}_\text{XY}=\left|D_{A_{1}}\phi\right|^{2}\qquad\leftrightarrow\qquad\mathcal{L}_\text{AH}=\left|D_{c}\Phi\right|^{2}-i\left[-\frac{1}{2\pi}c d A_1\right].
\end{align}
The mapping of the phases is such that positive mass deformations on one end maps to a negative deformation on the other end, $m_\Phi^2\leftrightarrow -m_\phi^2$.

In order to derive \eqref{eq:pv_dual} we start by taking the Abelian limit of Aharony's dualities, \eqref{eq:abelian_aharony}. In particular, take the $N = k = N_f =1$ and $N_s = 0$ limit of \eqref{eq:master_dual}, which yields the ``scalar + $U(1)_1$ $\leftrightarrow$ free fermion'' duality,
\begin{subequations}
\label{eq:scalar+flux}
\begin{align}
\mathcal{L}_{SU} & =i\bar{\psi}\Dslash_{A_1}\psi \\
\mathcal{L}_{U} & =\left|D_{c}\Phi\right|^{2}-i\left[\frac{1}{4\pi}cdc-\frac{1}{2\pi}c d A_1\right],
\end{align}
\end{subequations}
with $m_\psi\leftrightarrow -m_\Phi^2$. Meanwhile, the ``fermion + $U(1)_{-1/2}$ $\leftrightarrow$ WF scalar'' duality is obtained by taking the $N=k=N_s=1$ and $N_f=0$ limit,
\begin{subequations}
\label{eq:pvd_dual2}
\begin{align}
\mathcal{L}_{SU} & =\left|D_{A_{1}}\phi\right|^{2}-i\left[-\frac{1}{4\pi}A_1 d A_1\right],\label{eq:pvd xy}\\
\mathcal{L}_{U} & =i\bar{\Psi}\Dslash_{c}\Psi-i\left[\frac{1}{4\pi} cdc-\frac{1}{2\pi} c d A_1\right],\label{eq:pvd ferm+flux}
\end{align}
\end{subequations}
with $m_\phi^2\leftrightarrow m_\Psi$.

Deriving the bosonic particle-vortex duality from the above two dualities is straightforward. Note that we already have the XY model in \eqref{eq:pvd xy} up to the additional background Chern-Simons term. Hence, we should look for another bosonic theory dual to \eqref{eq:pvd ferm+flux}. To do so, add $-i\left[\frac{1}{4\pi}A_1 d A_1-\frac{1}{2\pi} A_1 d B_1\right]$ to each side of \eqref{eq:scalar+flux} and promote the $U(1)$ background field to be dynamical, $A_1\to a_1$. This gives the dual theories
\begin{subequations}
\label{eq:pv_dual3}
\begin{align}
\mathcal{L}_{SU}^\prime & =i\bar{\psi}\Dslash_{a_1}\psi -i\left[\frac{1}{4\pi}a_1 da_1-\frac{1}{2\pi}a_1 d B_1\right] \label{eq:pv_ferm2}\\
\mathcal{L}_{U}^\prime & =\left|D_{c}\Phi\right|^{2}-i\left[\frac{1}{4\pi}cdc-\frac{1}{2\pi}cda_1+\frac{1}{4\pi}a_1 d a_1-\frac{1}{2\pi}a_1 d B_1\right].\label{eq:pv_scalar3}
\end{align}
\end{subequations}
Since the action is quadratic in the newly promoted $a_1$ field we can integrate it out, which imposes the constraint $a_1 = c+B_1$. Plugging this in, we find
\begin{align}
\mathcal{L}_{U}^\prime =\left|D_{c}\Phi\right|^{2}-i\left[-\frac{1}{2\pi}cdB_1-\frac{1}{4\pi}B_1 d B_1\right].\label{eq:pvd_ah}
\end{align}
After relabeling the dynamical field in \eqref{eq:pv_ferm2} as $a_1\to c$ and changing the background field $B_1 \to A_1$, we see \eqref{eq:pv_ferm2} matches \eqref{eq:pvd ferm+flux}, and thus \eqref{eq:pvd_ah} is dual to \eqref{eq:pvd xy}. Canceling the common background Chern-Simons term, we arrive at the usual particle-vortex duality, \eqref{eq:pv_dual}. Note we get the relative mass flipping between the two ends of the duality since there is only a relative sign flip between $\psi$ and $\Phi$ masses.

We would now like to recast the derivation we just performed to motivate generalization to a two-node linear quiver. Fig. \ref{fig:pv_dual} schematically shows how we would like to view the derivation. Each of our dual theories in \eqref{eq:pvd_dual2} and \eqref{eq:pv_dual3} can be viewed as a two-node linear quiver, with the matter bifundamentally charged under the two nodes which it connects.

This is motivated by the fact that in Aharony's dualities \eqref{eq:aharony_schematic}, each matter field is fundamentally charged under both a dynamical gauge field and background global flavor symmetry. If we were to promote said flavor symmetry to be dynamical, the matter becomes a bifundamental and thus admits a natural description as a two-node quiver. This looks rather trivial since $SU(N)$ gauge groups for $N=1$ are nonsensical, but will generalize nicely for $N\geq 2$. For the Abelian case we will use $SU(1)$ as a placeholder for symmetries that can be gauged in the more general case.

To see this on the Abelian Higgs side, we will first shift the dynamical gauge field, $c\to c+a_1$, so that \eqref{eq:pv_scalar3} becomes
\begin{align}
\mathcal{L}_{U}^{\prime\prime} =\left|D_{c+a_1}\Phi\right|^{2}-i\left[\frac{1}{4\pi}c d c-\frac{1}{2\pi}a_1 dB_1\right].\label{eq:pvd_ah_alt}
\end{align}
In this form the scalar is bifundamentally charged under two $U(1)$ gauge groups, which represent the two nodes in the quiver theory. The dual to the Abelian Higgs model, \eqref{eq:pv_ferm2}, couples to a single dynamical $U(1)$ gauge field, $a_1$. This was previously the flavor symmetry but was promoted to a gauge symmetry in moving from \eqref{eq:scalar+flux} to \eqref{eq:pv_dual3}. As mentioned above, the gauge field belonging to the second node is absent only because we are working in the Abelian limit of Aharony's dualities. On the XY model end of the duality \eqref{eq:pvd xy}, $\phi$ couples to two $SU(1)$ fields, so it has no gauge couplings at all.\footnote{For our purposes here, we are ignoring the possibility of gauging the $U(1)$ global symmetry since its properties are well established in the particle-vortex duality as a global symmetry.}

The upshot of recasting the derivation in this form is that it readily generalizes to more complicated two-node quivers. One can use the more general Aharony's dualities to perform very similar steps as was done in the Abelian case. We'll see  particle-vortex duality generalizes to a duality of the form
\begin{multline}\label{eq:NonAbPV}
SU(N_1)_{-k_1}\times SU(N_2)_{-k_2} + \text{bifundamental scalar} \\
\qquad\leftrightarrow\qquad U(k_1)_{N_1-N_2}\times U(k_1+k_2)_{N_2} + \text{bifundamental scalar}.
\end{multline}
The bosonic particle-vortex duality is then just the $N_1=N_2=k_1=1$ and $k_2=0$ case. In Sec. \ref{sec:checks} we present further evidence of this interpretation by matching the spectrum of particles and vorticies in \eqref{eq:NonAbPV} in a manner similar to the Abelian case. Before we do this we demonstrate how we can systematically construct the non-Abelian quivers for an arbitrary number of nodes. This requires the use of the master duality when the number of nodes is greater than two.

\subsection{Building Non-Abelian Linear Quiver Dualities}
\label{sec:building_quivers}

Following the discussion in the previous subsection, our strategy in deriving dual descriptions of quiver gauge theories is to start with the master duality and gauge global symmetries on both sides of the duality in order to arrive at a duality for the resulting product gauge group. Since in a quiver gauge theory the gauge group associated with a given node sees the gauge groups associated with the neighboring nodes as global flavor symmetries, this roughly speaking amounts to dualizing the quiver one node at a time. While not a proof, this procedure suggests the resulting theories are dual. This basic idea had previously been pursued in ref. \cite{Jensen:2017dso} using Aharony's duality, but the flavor bounds put severe limitations on the quivers that were amendable to this analysis. In particular, the most interesting case with equal rank gauge groups on each node was out of reach. We will see that the master duality will help overcome many of these limitations.

To streamline the derivation it is helpful to follow ref. \cite{Jensen:2017dso} and rearrange BF terms to group the $SU(N)$ and $U(1)$ global symmetries together. Additionally, a key ingredient in matching this analysis to the existing particle-vortex duality will be the global $U(1)$ symmetries on either side of the duality. As such, we will be especially careful in keeping track of the global symmetries at every step.

We start by recalling how ref. \cite{Jensen:2017dso} derived their quiver transformations and generalize their method to the master duality. Starting from \eqref{eq:aharony_dual}, one can use the fact the $U(k)$ field can be separated into its Abelian and non-Abelian parts, i.e. $c=c^{\prime} +\tilde{c}\mathds{1}_{k}$, to perform a shift on the Abelian portion, $\tilde{c}\to\tilde{c}+\tilde{A}_{1}$. This allows one to rewrite $\mathcal{L}_{U}$ as
\begin{equation}
\mathcal{L}_{U}=i\bar{\Psi}\Dslash_{c+B+\tilde{A}_1}\Psi-i\left[\frac{N}{4\pi}\text{Tr}_{k}\left(cdc-i\frac{2}{3}c^{3}\right)-\frac{N k}{4\pi}\tilde{A}_{1}d\tilde{A}_{1}\right].
\end{equation}
Canceling the overall factor of $i\frac{N k}{4\pi}\tilde{A}_{1}d\tilde{A}_{1}$ on either side of the duality and defining the new $U(N_s)$ background field
$G_{\mu}\equiv B_{\mu}+\tilde{A}_{1\mu}\mathds{1}_{N_s}$, \eqref{eq:aharony_dual} becomes
\begin{subequations}
\begin{align}
\mathcal{L}_{SU} & = \left| D_{b^\prime+G}\phi \right|^2 -i\left[-\frac{k}{4\pi}\text{Tr}_{N}\left(b^{\prime}db^{\prime}-i\frac{2}{3}b^{\prime 3}\right)\right]\\
\mathcal{L}_{U} & =i\bar{\Psi}\Dslash_{c+G}\Psi-i\left[\frac{N}{4\pi}\text{Tr}_{k}\left(cdc-i\frac{2}{3}c^{3}\right)\right].
\end{align}
\end{subequations}
The procedure used in \cite{Jensen:2017dso} to derive new dualities is to promote the non-Abelian $U(N_s)$ global symmetry to be dynamical. Since both the $\phi$ and $\Psi$ matter is charged under $G$, this turns the matter into bifundamentals. Schematically, we denote the promoted duality as
\begin{equation}
SU(N)_{-k}\times U(N_s)_0 \qquad \leftrightarrow\qquad U(k)_{N-N_s/2}\times U(N_s)_{-k/2}.\label{eq:aharony dual 2 tw}
\end{equation}
This is subject to the flavor bound $N \geq N_s$.

In promoting the $U(1)$ global symmetry to a gauge symmetry, we get another $U(1)$ global symmetry which couples to the new gauge current on either side of the duality. If we wanted to make the coupling to the new background gauge field $\tilde{B}_1$ explicit, we would add a $-i\frac{1}{2\pi}\tilde{A}_1d\tilde{B}_1$ term to each side of the duality. This is completely analogous to the procedure preformed in ref. \cite{Karch:2016sxi}, where a new BF term was included with each promotion to represent the new $U(1)$-monopole symmetry on each side of the duality.

Below, we will sometimes apply this duality to strictly $SU$ gauge fields, in which case it is advantageous to only gauge the $SU$ part of the flavor symmetry, so that \eqref{eq:aharony dual 2 tw} becomes
\begin{equation}
SU(N)_{-k}\times SU(N_s)_0\qquad \leftrightarrow\qquad U(k)_{N-N_s/2}\times SU(N_s)_{-k/2}.\label{eq:aharony dual 4}
\end{equation}
Note that in this form of the duality each side retains the original global $U(1)$ symmetries and we do not obtain the additional global $U(1)$ as above.

We could apply the same procedures to the case where the $SU$ side contains the fermion and the $U$ side contains the scalar, where we would then find
\begin{align}
SU(N)_{-k+N_f/2}\times U(N_f)_{N/2}\qquad\leftrightarrow\qquad U(k)_{N}\times U(N_f)_0, \label{eq:aharony dual tw}
\end{align}
which matches the result found in ref. \cite{Jensen:2017dso} up to an overall shift in the level of the background term. This case is considered in more detail in Appendix \ref{appendix:duals}.


Now let us perform similar manipulations to the master duality in \eqref{eq:master_dual}. Since the Chern-Simons terms on the $U$ side are identical to \eqref{eq:mdb lu-1-1}, performing the same manipulations, \eqref{eq:master_LU} becomes
\begin{equation}
\mathcal{L}_{U}=\left|D_{c+\tilde{A}_{1}+C}\Phi\right|^{2}+i\bar{\Psi}\Dslash_{c+\tilde{A}_{1}+B+\tilde{A}_{2}}\Psi+\mathcal{L}_{\text{int}}^{\prime}-i\left[\frac{N_{1}}{4\pi}\text{Tr}_{k_{1}}\left(cdc-i\frac{2}{3}c^{3}\right)-\frac{N_{1}k_{1}}{4\pi}\tilde{A}_{1}d\tilde{A}_{1}\right].
\end{equation}
Again, we cancel the common $\tilde{A}_{1}$ Chern-Simons terms on either side of the duality. It will also be convenient to perform a shift to move the $\tilde{A}_2$ fields onto the $\psi$ and $\Phi$ matter, so we take $\tilde{A}_1\to\tilde{A}_1-\tilde{A}_2$ on either side of the duality. We could now combine the $U(1)$ and $SU(N_f)$ global symmetries into the definition of $E_{\mu}=C_{\mu}+\tilde{A}_{1\mu}\mathds{1}_{N_f}$ as we did in \eqref{eq:aharony dual 2 tw}. However, we will hold off on doing this since it is more convenient to keep the two global symmetries separate for our purposes. This leaves us with a duality of the form
\begin{subequations}
\begin{align}
\mathcal{L}_{SU} & =\left|D_{b^{\prime}+B+\tilde{A}_1}\phi\right|^{2}+i\bar{\psi}\Dslash_{b^{\prime}+ C + \tilde{A}_1-\tilde{A}_{2} }\psi+\mathcal{L}_{\text{int}}-i\left[\frac{N_f-k}{4\pi}\text{Tr}_{N}\left(b^{\prime}db^{\prime}-i\frac{2}{3}b^{\prime3}\right)\right]\nonumber \\
 & -i\left[\frac{N}{4\pi}\text{Tr}_{N_f}\left(CdC-i\frac{2}{3}C^{3}\right)+\frac{N N_f}{4\pi}(\tilde{A}_1-\tilde{A}_2)d(\tilde{A}_1-\tilde{A}_2)\right],\\
\mathcal{L}_{U} &= \left|D_{c+C + \tilde{A}_1-\tilde{A}_{2}}\Phi\right|^{2}+i\bar{\Psi}\Dslash_{c+B+\tilde{A}_1}\Psi+\mathcal{L}_{\text{int}}^{\prime}-i\left[\frac{N}{4\pi}\text{Tr}_{k}\left(cdc-i\frac{2}{3}c^{3}\right)\right].
\end{align}
\end{subequations}

At this point, we have two choices with how to treat the global symmetry associated with $\tilde{A}_2$. The first choice is to simply leave it as a global symmetry and gauge only the $SU(N_f)$ flavor symmetry associated with $C$.
In this form, each side of the master duality retains the $U(1)$ global symmetries associated with $\tilde{A}_1$ and $\tilde{A}_2$. Alternatively, we could also gauge the global symmetry associated with $\tilde{A}_2$. The latter of these cases will be useful for our purposes in this paper, so we define a $U(N_f)$ gauge field $G_{\mu}\equiv C_{\mu}-\tilde{A}_{2\mu}\mathds{1}_{N_f}$ to which $\psi$ and $\Phi$ couple. After gauging the $U(N_f)$ and $SU(N_s)$ global symmetries, this leaves us with
\begin{equation}
SU(N)_{-k+N_f/2}\times U(N_f)_{N/2}\times SU(N_s)_0 \quad\leftrightarrow\quad U(k)_{N-N_s/2}\times U(N_f)_0 \times SU(N_s)_{-k/2}.\label{eq:master dual tw}
\end{equation}
Similar to Aharony's duality, in gauging the $U(N_f)$ symmetry which is associated with $G$, we pick up an additional monopole $U(1)$ symmetry on either side of the duality which couples to the newly gauged $\tilde{A}_2$ field. We will denote the background gauge field associated with said symmetry by $\tilde{B}_2$. For completeness, we consider the master duality with all global symmetries gauged in Appendix \ref{appendix:duals}.

Note that we can modify either of the above dualities by adding additional background flavor levels to either side of the duality before promotion. As a reminder, these dualities are subject to the flavor bound $k\geq N_f $ and $N\geq N_s$, but $\left(k,N \right)\ne\left(N_f,N_s\right)$. In the $N_f=0$ and $N_s=0$ limits, \eqref{eq:master dual tw} reduces to \eqref{eq:aharony dual 4} and \eqref{eq:aharony dual tw}, respectively, with appropriate relabeling.

\subsubsection*{Four-Node Example}

\begin{figure}
\begin{centering}
\includegraphics[scale=0.6]{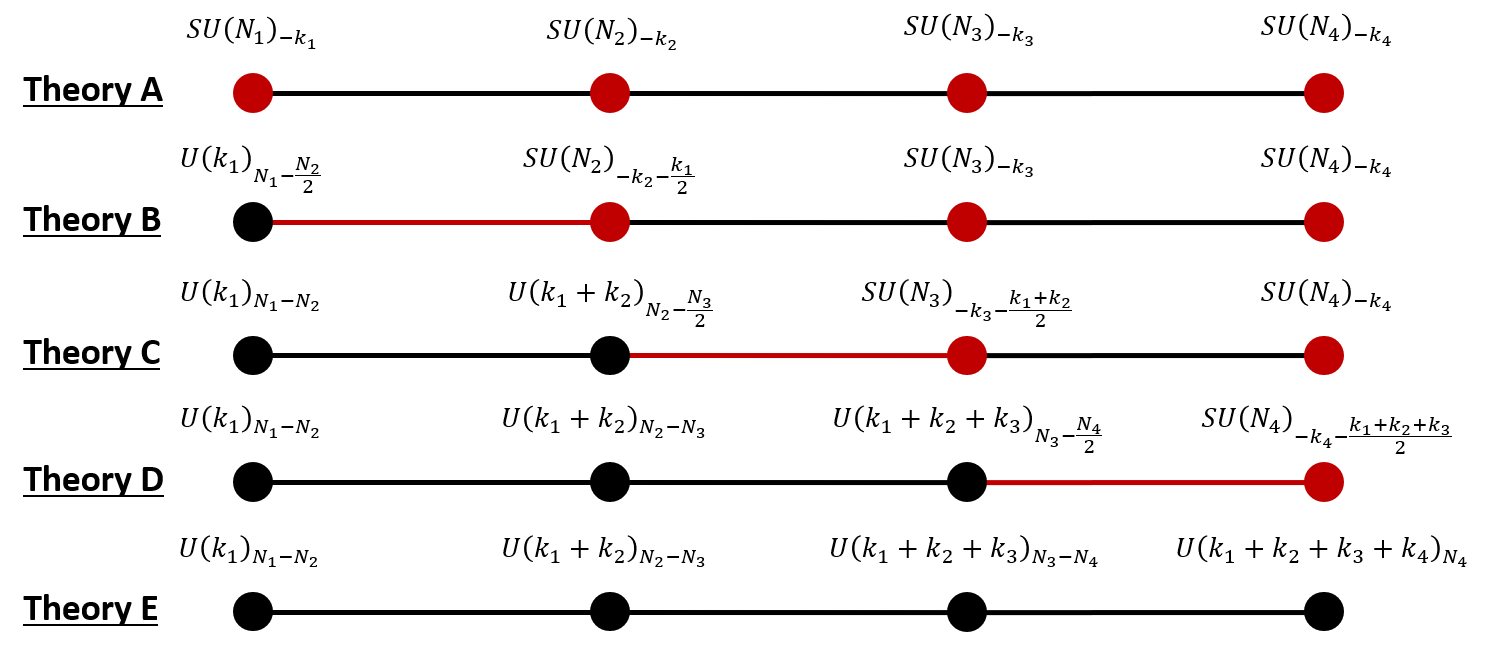}
\par\end{centering}
\caption{Dualizing a linear quiver. Red nodes are $SU$ gauge groups and black nodes are $U$ groups. Black (red) links are bifundamental bosons (fermions).  Applying Aharony's duality to the leftmost link turns the scalar into a fermion. Then, applying the master duality repeatedly moves said fermion across the quiver until it reaches the final link where Aharony's duality can again be used to turn the fermion into a boson. \label{fig:linear quiver tw}}
\end{figure}

Let us now use the dualities we've defined to dualize a four-node quiver. Walking through this construction will make generalization to the $n$-node case straightforward. We begin with the $SU$ side of the theory
\begin{equation}
\text{Theory A:}\qquad SU\left(N_{1}\right)_{-k_{1}}\times SU\left(N_{2}\right)_{-k_{2}}\times SU\left(N_{3}\right)_{-k_{3}}\times SU\left(N_{4}\right)_{-k_{4}}.\label{eq:linear SU4 tw}
\end{equation}
This theory has the bifundamental scalars which have charges as given in Table \ref{tab:Charges-of-the bifund linear tw}. In what follows we will assume that $N_{1}\geq N_{2}\geq N_{3}\geq N_{4}$ as well as $k_{i}\geq0$ for $i=1,2,3,4$. Although this is not the most general case, below we will find this is required to avoid flavor bounds to get to the desired $U$ theory. Each of the bifundamentals is charged under a global $U(1)$ symmetry which rotates its overall phase, giving this side of the duality a $[U(1)]^3$ global symmetry.

We will denote the bifundamental scalars living between nodes $j$ and $j+1$ by $Y_{j,j+1}$ and $X_{j,j+1}$ on the $SU$ and $U$ side of the duality, respectively. The masses of the $U$ bifundamentals are denoted by $m_{j,j+1}$ while we will use $M_{j,j+1}$ for those on the $SU$ side.

\begin{table}
\begin{centering}
\begin{tabular}{|c|c|c|c|c|}
\hline
\textbf{Theory A} & $SU(N_1)_{-k_1}$ & $SU(N_2)_{-k_{2}}$ & $SU(N_3)_{-k_{3}}$ & $SU(N_4)_{-k_{4}}$\tabularnewline
\hline
$Y_{1,2}$ & $\square$ & $\square$ & $1$ & $1$\tabularnewline
\hline
$Y_{2,3}$ & $1$ & $\square$ & $\square$ & $1$\tabularnewline
\hline
$Y_{3,4}$ & $1$ & $1$ & $\square$ & $\square$\tabularnewline
\hline
\end{tabular}
\par\end{centering}
\caption{Charges of the bifundamental matter in our linear quivers. $\square$ denotes the matter transforms in the fundamental representation of the corresponding gauge group. \label{tab:Charges-of-the bifund linear tw}}
\end{table}

Before embarking on deriving the duality, note that the uniform mass deformations of (\ref{eq:linear SU4 tw}) are given by
\begin{subequations}
\label{eq:su_deforms}
\begin{align}
(\text{A1})\qquad M_{i,i+1}^{2}>0:\qquad & SU\left(N_{1}\right)_{-k_{1}}\times SU\left(N_{2}\right)_{-k_{2}}\times SU\left(N_{3}\right)_{-k_{3}}\times SU\left(N_{4}\right)_{-k_{4}}\\
(\text{A2})\qquad M_{i,i+1}^{2}<0:\qquad & SU\left(N_{1}-N_{2}\right)_{-k_{1}}\times SU\left(N_{2}-N_{3}\right)_{-k_{1}-k_{2}}\times SU\left(N_{3}-N_{4}\right)_{-k_{1}-k_{2}-k_{3}}\nonumber \\
 & \times SU\left(N_{4}\right)_{-k_{1}-k_{2}-k_{3}-k_{4}}.
\end{align}
\end{subequations}
Here we have been careful to account for which gauge group is Higgsed by each bifundamental scalar. The bifundamental scalar $Y_{i,i+1}$ has $N_{i}\times N_{i+1}$ components. Below we will always view the smaller of the two gauge groups to be associated with the ``flavor'' symmetry of the bifundamentals. As such, if we assume the Higgsing to be maximal, the Higgsing can be thought of as acting on the ``color'' gauge group, i.e. the group with the larger rank, while leaving the flavor group unchanged.\footnote{For example, for a bifundamental coupled to $SU(N_1)_{c}$ and $SU(N_2)_{f}$ with $N_{1}\geq N_{2}$, what occurs can be best understood by first splitting  $ SU(N_1)_c \times SU(N_2)_f \to SU(N_1-N_2)_c\times SU(N_2)_c\times SU(N_2)_f$. Since the bifundamental is maximally Higgsed in the $SU(N_2)$ subgroup, the unbroken part of the two $SU(N_2)$ factors is their diagonal, leaving $SU(N_1-N_2)_c\times SU\left(N_{2}\right)_{\text{diag}}$. Thus, saying the flavor group is unchanged is a merely a convenient relabeling. Also note that the Chern-Simons level of the new flavor group will be the sum of the original flavor and color Chern-Simons levels.} Since to meet flavor bounds below we have assumed $N_{1}\geq N_{2}\geq N_{3}\geq N_{4}$, this means vacuum expectation value takes the form
\begin{equation}
\langle Y_{i,i+1}^{a_{i}a_{i+1}} \rangle\propto\left(\begin{array}{c}
\mathds{1}_{N_{i+1}}\\
0
\end{array}\right)^{a_{i}a_{i+1}}\label{eq:su vev tm}
\end{equation}
with $a_i,b_i$ and $a_{i+1}, b_{i+1}$ gauge indices to $SU(N_i)$ and $SU(N_{i+1})$, respectively. Reassuringly, no gauge group acquires a negative rank with this Higgsing pattern.

Returning to our derivation of the non-Abelian linear quiver dualities, we will now show five theories are dual to one another,
\begin{equation}
\text{Theory A}\quad\leftrightarrow\quad\text{Theory B}\quad\leftrightarrow\quad\cdots\quad\leftrightarrow\quad\text{Theory E},
\end{equation}
by sequentially dualizing each node from left to right, see Fig. \ref{fig:linear quiver tw}. To begin, we apply Aharony's duality \eqref{eq:aharony dual 4} to the first node to obtain
\begin{equation}
\text{Theory B:}\qquad U\left(k_{1}\right)_{N_{1}-N_{2}+\frac{N_{2}}{2}}\times SU\left(N_{2}\right)_{-k_{2}-k_{1}+\frac{k_{1}}{2}}\times SU\left(N_{3}\right)_{-k_{3}}\times SU\left(N_{4}\right)_{-k_{4}}.
\end{equation}
Flavor bounds are satisfied so long as $N_{1}\geq N_{2}$. This turns the bifundamental scalar on the link between nodes one and two into a bifundamental fermion.  The $U(1)$ global symmetry of the first bifundamental becomes a monopole symmetry for the Abelian part of the gauge field which lives on the first node. This will be a common theme as we sequentially step through nodes and the details are shown in Appendix \ref{appendix:sym}.

Dualizing the $SU(N_2)$ node is where we will need to use something new. We could try applying Aharony's dualities \eqref{eq:aharony dual 2 tw} or \eqref{eq:aharony dual tw} to the second node. However, one will inevitably run into flavor bound issues since nodes with links on two sides require a $SU(N_{i-1}+N_{i+1})$ flavor symmetry, which exceeds the $SU(N_i)$ color symmetry for the cases we are interested in here.

Notice that since the master duality has two types of matter it has two \emph{separate }flavor symmetries, each subject to its own flavor bound. This is useful for dualizing the nodes with two links and, furthermore, has the correct matter content since node two in Theory B has both a bifundamental scalar and fermion attached to it. However, the master duality is quite a bit different from Aharony's original dualities in that it requires additional interactions terms between the scalars and fermions on a given side of the duality, as given in \eqref{eq:master_ints}. Let us consider how we could introduce such interactions terms and how they affect the theories we are considering.

\subsubsection*{Including Bifundamental Interactions}

The interaction we need in order to apply the master duality in theory B is
\begin{align}
\text{Theory B:}\qquad & C_{2}^{(B)}\left(\bar{\psi}_{1,2}^{a_{2}a_{1}}Y_{a_{2}a_{3}}^{2,3}\right)\left(Y_{2,3}^{\dagger b_{2}b_{3}}\psi_{b_{2}a_{1}}^{1,2}\right).\label{eq:master matter int B tw}
\end{align}
Here $C_{2}^{(B)}$ is the coefficient of the interaction on the second node in theory B and we have not yet committed to its sign or magnitude, but we will do so later by matching TFTs. In what follows, it will be useful to associate each interaction term with one of the interior nodes of the quiver.

In order to give rise to \eqref{eq:master matter int B tw}, we must backtrack slightly since a similar interaction should also then be present in theory A in its dualized form. The exact matching of the interactions between theories A and B is quite subtle and requires auxiliary field techniques that were originally introduced in the large $N$ and $k$ literature \cite{Minwalla:2015sca}. Here we will only give a schematic overview. The full details of this matching are given in Appendix \ref{appendix:int}.

Recall that the purpose of the interaction term in the master duality is to ensure that when the scalars acquire a vacuum expectation value we also gain an additional mass term for the fermions \cite{Benini:2017aed,Jensen:2017bjo}. This was vital for matching the phases and TFTs on either side of the duality. Importantly, regardless of the sign of the mass deformation, the fermions never condense and thus there is no opportunity for a fermion condensate to influence the mass of the scalars through the same interaction term. If the fermions did condense, this would yield a very different looking phase diagram than that found in refs. \cite{Benini:2017aed,Jensen:2017bjo}.

Here we identify the fermions with scalars, which \emph{can} condense when their quadratic term goes negative. In order to match the TFT phase diagrams of theories A and B under mass deformations, we need to make sure the interaction term does not allow the $Y_{1,2}$ condensate to influence the $Y_{2,3}$ mass. In Appendix \ref{appendix:int} we derive an interaction term which has the desired properties: this term will cause a nonzero vacuum expectation value for the $Y_{2,3}$ bifundamental to give a positive or negative mass to $Y_{1,2}$, depending on the sign of the coefficients $C_{2}^{(A)}$. However, the opposite effect cannot occur: $Y_{1,2}$ acquiring a vacuum expectation value \emph{cannot} influence the mass of $Y_{2,3}$. More generally, for the $SU$ side of the duality, the vacuum expectation value of a link can only affect nodes/links to its \emph{left}. The interaction term is unidirectional
as it is for the original master duality.

We will \emph{schematically} denote the interactions we add to Theory A as
\begin{align}
\text{Theory A:}\qquad & C_{2}^{\left(A\right)}\left(Y_{1,2}^{\dagger a_{2}a_{1}}Y_{a_{2}a_{3}}^{2,3}\right)\left(Y_{2,3}^{\dagger b_{2}a_{3}}Y_{b_{2}a_{1}}^{1,2}\right)\label{eq:master matter int A tw}
\end{align}
with the understanding that the true interaction is as given in \eqref{eq:su_ints}. Eq. \eqref{eq:master matter int A tw} is equivalent to \eqref{eq:su_ints} if we simply ignore the fact that when the $Y_{1,2}$ acquires a vacuum expectation value the interaction term gives a mass to $Y_{2,3}$, so we will do so henceforth for brevity. An analogous interaction term is added to node three as well since it will be needed when stepping from theory C to D.

Having introduced the necessary interaction term, we subsequently apply the master duality \eqref{eq:master dual tw} to the second and third nodes, this gives
\begin{align}
\text{Theory C:}\qquad & U(k_1)_{N_{1}-N_{2}}\times U(k_{1}+k_{2})_{N_{2}-\frac{N_{3}}{2}}\times SU(N_{3})_{-k_{3}-k_{1}-k_{2}+\frac{k_{1}+k_{2}}{2}}\times SU(N_4)_{-k_{4}}\\
\text{Theory D:}\qquad & U(k_1)_{N_{1}-N_{2}}\times U(k_{1}+k_{2})_{N_{2}-N_{3}}\times U(k_{1}+k_{2}+k_{3})_{N_{3}-\frac{N_{4}}{2}}\nonumber \\
 & \times SU(N_4)_{-k_{4}-k_{1}-k_{2}-k_{3}+\frac{k_{1}+k_{2}+k_{3}}{2}}.
\end{align}
These in turn require the flavor bounds $N_{2}\geq N_{3}, k_{2}\geq 0$ and $N_{3}\geq N_{4}, k_{3}\geq 0$, respectively.\footnote{More precisely, this should exclude the double saturation cases where $N_2=N_3$ and $k_2=0$ or $N_3=N_4$ and $k_3 =0$. We will not make note of such special cases henceforth since they will not be relevant for our purposes.} Each application of the master duality changes a boson link to a fermion link and vice versa, effectively driving the single fermion link down the quiver, see Fig. \ref{fig:linear quiver tw}. As with the duality relating theories A and B, the application of the master duality above changes the global $U(1)$ symmetry across the duality. Specifically, it changes the $U(1)$ global symmetry under which $Y_{2,3}$ was charged to a monopole-like symmetry which couples to the Abelian part of gauge field on the second node. A completely analogous transformation occurs for the baryon number symmetry of $Y_{3,4}$. The details of how this occurs are shown in Appendix \ref{appendix:sym}.

Finally, to arrive at the desired dual theory we again apply Aharony's duality \eqref{eq:aharony dual tw} to the last node.
Flavor bounds require $k_{4}\geq0$. This ultimately gives
\begin{equation}
\text{Theory E:}\qquad U\left(k_{1}\right)_{N_{1}-N_{2}}\times U\left(k_{1}+k_{2}\right)_{N_{2}-N_{3}}\times U\left(k_{1}+k_{2}+k_{3}\right)_{N_{3}-N_{4}}\times U\left(k_{1}+k_{2}+k_{3}+k_{4}\right)_{N_{4}}.
\end{equation}
Note that the fourth node does not pick up a monopole-like global symmetry for its Abelian gauge field. This is related to the fact we have one more node than bifundamentals and is also a feature of the dualities in ref. \cite{Jensen:2017dso} and ABJM theory \cite{Aharony:2008ug}.

Following the mass identifications through the dualities, we see $M_{i,i+1}^2\leftrightarrow -m_{i,i+1}^2$. The uniform mass deformations of Theory E are
\begin{subequations}
\begin{align}
(\text{E1})\qquad m_{i,i+1}^{2}<0:\qquad & U\left(k_{1}\right)_{N_{1}}\times U\left(k_{2}\right)_{k_{2}}\times U\left(k_{3}\right)_{N_{3}}\times U\left(k_{4}\right)_{N_{4}}\\
(\text{E2})\qquad m_{i,i+1}^{2}>0:\qquad & U\left(k_{1}\right)_{N_{1}-N_{2}}\times U\left(k_{1}+k_{2}\right)_{N_{2}-N_{3}}\times U\left(k_{1}+k_{2}+k_{3}\right)_{N_{3}-N_{4}} \nonumber \\
& \times U(k_{1}+k_{2}+k_{3}+k_{4})_{N_{4}}
\end{align}
\end{subequations}
which, reassuringly, are level-rank dual to the phases considered above in \eqref{eq:su_deforms}. Here once again some care is required for the Higgs phase. Since we are assuming all $k_{i}\geq0$ in order to meet the flavor bounds above, the maximal Higgsing vacuum expectation value is, using block matrix notation,
\begin{equation}
\langle X_{i,i+1}^{a_{i}a_{i+1}}\rangle\propto\left(\begin{array}{cc}
\mathds{1}_{K_{i}} & 0\end{array}\right)^{a_{i}a_{i+1}}.\label{eq:u vev tw}
\end{equation}
where $a_{i},a_{i+1}$ are gauge indices to $U(K_{i})$ and $U(K_{i+1})$, respectively and we have defined the shorthand
\begin{equation}
K_{j}\equiv\sum_{i=1}^{j}k_{i}.\label{eq:big K def tw}
\end{equation}

Of course, as we apply all the aforementioned dualities the matter interaction terms are changing as well. We also end up with the interaction term \eqref{eq:u_ints} between adjacent bifundamental scalars for theory E. The interaction is such that a bifundamental scalar vacuum expectation can only affect nodes/links to its \emph{right} now. \footnote{This is most obvious to see when moving from Theory C to Theory D. In theory C, the $X_{1,2}$ bifundamental can influence the mass of the fermion on the $2,3$ link, but not vice versa. Hence, in Theory D the interaction between $X_{1,2}$ and $X_{2,3}$ should obey the same rule to get a matching of TFTs.} The analog of \eqref{eq:master matter int A tw}, which is a schematic stand-in for \eqref{eq:u_ints}, is
\begin{equation}
\text{Theory E:}\qquad C_{2}^{\left(E\right)}\left(X_{1,2}^{\dagger a_{2}a_{1}}X_{a_{2}a_{3}}^{2,3}\right)\left(X_{2,3}^{\dagger b_{2}a_{3}}X_{b_{2}a_{1}}^{1,2}\right).\label{eq:U theory ints tw}
\end{equation}
where now we ignore the fact that when $X_{2,3}$ acquires a vacuum expectation value it gives a mass to $X_{1,2}$.

\subsubsection*{Effect of Interactions}

We would now like to show that these interaction terms are vital for a matching the mass deformed TFTs. Although we found a matching between phases for the completely gapped/Higgsed phases above, these were very special cases. In order to observe the expected partial gapping/Higgsing behavior to apply to theta walls, we need to carefully treat the interactions.

First it will be helpful to specialize to a particular sign and magnitude of interaction terms coefficients, $C_I^{(A)}$ and $C_I^{(E)}$ for $I=2,3$ (i.e. all internal nodes). Specifically for the purposes of matching onto the phases of \eqref{eq:su_quiver_ym}, we take $C_{I}^{(A)}<0$ and $C_{I}^{(E)}<0$.
\footnote{We must choose the coefficients of the interactions to be the same sign for a matching of TFTs. To see this, first note that we use the master duality once on each internal node, and under the master duality the interaction term flips sign \cite{Benini:2017aed,Jensen:2017bjo}. An additional sign flip comes from the application of the master duality for the node to the left of an internal node, which sign flip when changing the fermions in the interaction to bosons.} We also assume $|C_{I}^{(E)}|\to \infty$ such that $|C_{I}^{(E)}|\gg|m_{i,i+1}^{2}|$. Although not considered in \cite{Benini:2017aed,Jensen:2017bjo}, it is straightforward to check that a very large interaction term on one side of the master duality implies a very small interaction term on the other side of the duality.\footnote{To see this, let us specialize to the notation used in \cite{Jensen:2017bjo}. Here we saw that by changing the sign of $c_{4}$ and $c_{4}^{\prime}$, one could change the location of the ``singlet critical line''. Smoothly changing $c_{4}\to-c_{4}$ causes the line to move from phase IV to phase III. Meanwhile, changing $c_{4}^{\prime}\to-c_{4}^{\prime}$ to move from phase IV' to phase I'. Thus, for example, we can shrink the size of region IVb and IVb' by decreasing the magnitude of $c_{4}^{\prime}$ and increasing the magnitude of $c_{4}$. In the limit $c_{4}^{\prime}\to0$, we must take $\left|c_{4}\right|\to\infty$. } Hence,  $C_{I}^{(A)}\to 0$ so the hierarchy $M_{i,i+1}^{2}\gg C_{I}^{(A)}>0$ holds for all mass deformations. In such a limit we can effectively ignore the interaction terms on the $SU$ side of the duality.

The choice of magnitudes above has the added effect of simplifying the analysis of the interaction terms and the TFT structure. It is possible to derive quiver theories for more general interaction coefficients and still find matching TFTs, but we leave such analysis for future work.

Let us now consider the effect of the interaction terms on the $U$ side of the duality. The two interaction terms of theory E are given by
\begin{equation}
C_{2}^{\left(E\right)}\left(X_{1,2}^{\dagger a_{2}a_{1}}X_{a_{2}a_{3}}^{2,3}\right)\left(X_{2,3}^{\dagger b_{2}a_{3}}X_{b_{2}a_{1}}^{1,2}\right)+C_{3}^{\left(E\right)}\left(X_{2,3}^{\dagger a_{3}a_{2}}X_{a_{3}a_{4}}^{3,4}\right)\left(X_{3,4}^{\dagger b_{3}a_{4}}X_{b_{3}a_{2}}^{2,3}\right).\label{eq:four node ints tw}
\end{equation}
Consider the case when $X_{1,2}$ acquires a nonzero vacuum expectation value as in \eqref{eq:u vev tw}. This term breaks $U(K_2)$ down to $U(K_{2}-K_{1})$, which is the usual effect of the Higgsing. Additionally, the first interaction term \eqref{eq:four node ints tw} becomes
\begin{equation}
-\left(X_{1,2}^{\dagger a_{2}a_{1}}X_{a_{2}a_{3}}^{2,3}\right)\left(X_{2,3}^{\dagger b_{2}a_{3}}X_{b_{2}a_{1}}^{1,2}\right)\propto-\left(\begin{array}{cc}
\mathds{1}_{K_{1}} & 0\\
0 & 0
\end{array}\right)_{b_{2}}^{a_{2}}X_{2,3}^{\dagger b_{2}a_{3}}X_{a_{2}a_{3}}^{2,3}.\label{eq:x12 breaking tw}
\end{equation}
Hence the vacuum expectation value of $X_{1,2}$ shows up as a negative mass deformation for the first $K_{1}$ components of $X_{2,3}$ and thus \emph{also} breaks the $U(K_{3})$ to $U(K_{3}-K_{1})$. As such, except for the $X_{1,2}$ link, each bifundamental can acquire a mass deformation from two different sources: its explicit mass term as well as the interaction terms to its left. As an example, let us  assume $m^2_{2,3}=m^2_{3,4}=0$ but $m^2_{1,2}< 0$. Then $X_{2,3}$ acquires a vacuum expectation value from \eqref{eq:x12 breaking tw}. The interaction term between $X_{2,3}$ and $X_{3,4}$ also means $X_{3,4}$ gets a negative mass shift,
\begin{equation}
-\left(X_{2,3}^{\dagger a_{3}a_{2}}X_{a_{3}a_{4}}^{3,4}\right)\left(X_{3,4}^{\dagger b_{3}a_{4}}X_{b_{3}a_{2}}^{2,3}\right)\propto-\left(\begin{array}{cc}
\mathds{1}_{K_{1}} & 0\\
0 & 0
\end{array}\right)_{b_{3}}^{a_{3}}X_{3,4}^{\dagger b_{3}a_{4}}X_{a_{3}a_{4}}^{3,4}.\label{eq:x23 x34 int tw}
\end{equation}
breaking $U(K_{4})$ to $U(K_{4}-K_{1})$ and giving the first $K_{1}$ components an additional negative mass deformation. If there are no other mass term deformations, this effect cascades to the \emph{right} across the entire quiver.

Now let us consider how this changes if $X_{2,3}$ also had a mass deformation. A negative mass deformation would further break down the $U(K_{3})$ subgroup to $U(K_{3}-K_{2})$, and this could also propagate down the quiver, as we just discussed. Positive mass deformations are slightly more tricky since we need to consider them in two regimes. First consider the case where the negative mass deformation from \eqref{eq:x12 breaking tw} is larger than that of the mass term for $X_{2,3}$. Then the results considered above are unchanged, $X_{1,2}$ is still partially broken. When the mass term has a larger positive mass contribution than that of \eqref{eq:x12 breaking tw}, $X_{2,3}$ is completely gapped. This means none of the components have a nonzero vacuum expectation value, and thus the interaction \eqref{eq:x23 x34 int tw} contributes no mass to $X_{3,4}$. In other words, if the mass term for $X_{2,3}$ is large enough, it can stop of propagation of $X_{1,2}$'s breaking down the quiver. We have avoided this case by assuming $\left|C_{I}^{\left(E\right)}\right|\gg\left|m_{i,i+1}^{2}\right|$, thereby forbidding large positive mass deformations from blocking the propagation of the breaking down the quiver.

What about when $X_{2,3}$ acquires a negative mass deformation but the $X_{1,2}$ and $X_{3,4}$ mass terms are untouched? By the same reasoning above, $X_{3,4}$ will also acquire a negative mass shift to its first $K_{2}$ components via the interaction terms, causing the breaking of $U(K_3)$ and $U(K_4)$ to $U(K_3-K_2)$ and $U(K_4-K_2)$, respectively. However, as we have been careful to argue in Appendix \ref{appendix:int}, $X_{2,3}$'s vacuum expectation value should not be able to influence nodes/links to its left. Hence $X_{1,2}$ is unaffected.


We are now in the position to consider mass deformations which are partially Higgsed. That is, not all bifundamental masses are taken to be the same sign. Specifically, consider the case where the bifundamentals $Y_{1,2}$ and $Y_{3,4}$ are Higgsed and $Y_{2,3}$ is gapped (corresponding via mass identifications to $X_{1,2}$ and $X_{3,4}$ being gapped and $X_{2,3}$ being Higgsed). This yields the phases
\begin{subequations}
\begin{align}
(\text{A3})\quad:\quad & SU(N_{1}-N_{2})_{-k_{1}}\times SU(N_{2})_{-k_{1}-k_{2}}\times SU(N_{3}-N_{4})_{-k_{3}}\times SU(N_{4})_{-k_{3}-k_{4}}\label{eq: partial Higgs A tw}\\
(\text{E3})\quad:\quad & U(k_1)_{N_{1}-N_{2}}\times U(k_{1}+k_{2})_{N_{2}}\times U(k_3)_{N_{3}-N_{4}}\times U(k_{3}+k_{4})_{N_{4}}\label{eq:partial Higgs B tw}
\end{align}
\end{subequations}
which are level-rank dual to one another! Note that the interactions are vital for us to reach this conclusion. We have used the fact that $X_{2,3}$ acquiring a vacuum expectation value breaks the $U(K_2)$ subgroup of $U(K_3)\to U(k_3)_{N_{3}-N_{4}}\times U(K_2)_{N_{3}-N_{4}}$ and also, due to the interaction term, $U(K_4)_{N_{4}}\to U(k_3+k_4)_{N_{4}}\times U(K_2)_{N_{4}}$. Without such terms we would have found the $U(K_{4})_{N_{4}}$ group unbroken, yielding the TFT
\begin{equation}
(\text{E}\bar{\text{3}})\qquad:\qquad U(k_1)_{N_{1}-N_{2}}\times U(k_{1}+k_2)_{N_{2}-N_{4}}\times U(k_3)_{N_{3}-N_{4}}\times U(K_4)_{N_{4}}
\end{equation}
which is clearly not level-rank dual to \eqref{eq: partial Higgs A tw}.

\subsubsection*{Generalization to $n$ Nodes}

Now let us generalize this prescription to an arbitrary number of nodes. For $n$ nodes, there is a duality between the following two theories:
\begin{subequations}
\label{eq:general_n}
\begin{align}
\text{Theory A:}\qquad & SU(N_{1})_{-k_{1}}\times\prod_{i=2}^{n}\left[SU(N_{i})_{-k_{i}}\times\text{bifundamental }Y_{i-1,i}\right]\\
\text{Theory B:}\qquad & \prod_{i=1}^{n-1}\left[U(K_{i})_{N_{i}-N_{i+1}}\times\text{bifundamental }X_{i,i+1}\right]\times U(K_{n})_{N_{n}}
\end{align}
\end{subequations}
where flavor bounds require $k_i\geq 0$ and $N_1 \geq N_2 \geq \ldots \geq N_n$. As with the above case, these theories can be shown to be dual by systematically applying Aharony's duality \eqref{eq:u ferm tw} to the first node, the master duality to every two-link node, and then Aharony's other duality \eqref{eq:u scalar tw} to the last node. The master duality is not needed for the two-node case.

Implied above are interaction terms on the $U$ side of the duality of the form of \eqref{eq:u_ints}. Equivalently, we can use the schematic interaction
\begin{align}
\mathcal{L}^{(B)}\supset & \sum_{I=2}^{n-1}C_{I}^{(B)}\left(X_{I-1,I}^{\dagger a_{I}a_{I-1}}X_{a_{I}a_{I+1}}^{I,I+1}\right)\left(X_{I,I+1}^{\dagger b_{I}a_{I+1}}X_{b_{I}a_{I-1}}^{I-1,I}\right)
\end{align}
with the understanding that such interactions can only give mass terms to the link on their left. Here, $a_{i}$, $b_{i}$ the gauge indices of the $i$th node and $C_{I}^{(B)} \to -\infty$ so that $|C_{I}^{(B)}|\gg m^2_{i,i+1}$.  In this limit, on the $SU$ side of the duality the interaction terms are very small and have no effect on the mass deformed phases, so we ignore them.

To summarize the interaction behavior on the $U$ side: a bifundamental scalar $X_{j,j+1}$ acquiring a nonzero vacuum expectation value affects nodes/links to the right but \emph{not }to the left. Namely, it causes all bifundamental scalars (i.e $X_{i,i+1}$ with $i>j$) to acquire a similar vacuum expectation for the first $K_{j}$ components. This in turn causes a breaking of all gauge groups nodes $i>j$ to $U(K_{i}-K_{j})$. Note this effect can compound, so if $X_{j,j+1}$ and $X_{\ell,\ell+1}$ acquire a vacuum expectation value from their respective mass deformations, the gauge group on node $i>j>\ell$ undergoes breaking $U(K_i)\to U(K_i- K_j) \times U(K_j-K_\ell) \times U(K_\ell)$.

As mentioned earlier, other dualities which flow to other TFTs can be constructed by changing the sign/magnitude of the interaction terms, but such considerations are left for future work.

\subsection{Self-Consistency Checks}\label{sec:checks}

Returning to the bosonic particle-vortex duality, it should now be clear the derivation we outlined in Sec. \ref{sec:boson_pv} is the two-node case of the more general non-Abelian linear quivers with values
\begin{align}
N_1=N_2=k_1 & = 1,\qquad k_2 = 0,
\end{align}
which is shown in Fig. \ref{fig:pv_dual}. Note this saturates all flavor bounds and carries the minimum value of parameters without being completely trivial, so the particle-vortex duality can be thought of as the simplest case of an infinite class of $2+1$ dimensional Bose-Bose dualities. Additionally, it is clear that the derivation of the two-node quiver requires no master duality since there are no nodes connected to two links.

Another helpful tool in analyzing the more general non-Abelian quiver dualities as well as comparing them to the holographic dualities in Sec. \ref{sec:jensen} will be comparing the spectrum of the two theories. To this end, let us briefly review how the spectra of the particle-vortex duality match on either side of the duality.

First consider the case when $\Phi$ acquires a vacuum expectation value in \eqref{eq:pvd_ah} through a negative mass deformation. It is well known the breaking of the $U(1)$ gauge symmetry gives rise to vortex solutions of finite mass charged under $B_1$ flux. Since there is no dynamical Chern-Simons term on this end, there is no funny business with flux attachment or alternative vortex solutions. These vortices carry flux charge under the broken $U(1)$ gauge group which can be seen by looking at the asymptotic behavior of the gauge field.

Now consider the Abelian-Higgs model but instead in the form which is more amenable to matching onto the non-Abelian quivers (i.e. \eqref{eq:pvd_ah_alt}). In this form we have two $U(1)$ gauge fields, one of which is redundant and can be integrated out. When $\langle\Phi\rangle \sim v$, it forces the breaking of $U(1) \times U(1)\to U(1)_{A}$, where $U(1)_{A}$ is the subgroup where the two $U(1)$ transformations act oppositely on $\Phi$, leaving it invariant. Again, the breaking of a $U(1)$ symmetry ensures that there are vortex solutions which are charged under the flux of the broken symmetry. In this case, it corresponds to a nonzero winding of both $a_1$ and $c$ at spatial infinity, since the broken $U(1)$ group is where they are set equal to one another (i.e. $U(1)_\text{diag}$). For the vortex solution where $a_1=c$ energy contributions from the Chern-Simons terms drop out, as they should since they weren't present in \eqref{eq:pvd_ah}. Since there is nonzero $a_1$ flux the vortex is charged under the background $B_1$ field. Also note that the vortex has finite mass proportional to the vacuum expectation value of the scalar. As expected, we reach the same conclusions when working from \eqref{eq:pvd_ah}, albeit in a slightly more complicated manner.

Due to the mass identification, the phase where $\Phi$ has a vacuum expectation value should be identified with the phase where $\phi$ is simply gapped. The $U(1)$ global symmetry is unbroken and $\phi$ excitations of \eqref{eq:pvd xy} are charged under the $B_1$ field, which are identified with the vortices on the opposite side of the duality.

Meanwhile, for mass deformations where $\phi$ obtains a vacuum expectation value and $\Phi$ is gapped, the $U(1)$ global symmetry on both sides of the duality is broken. This is straightforward to see on the $\phi$ side of the duality and is made clear on the $\Phi$ side by rewriting the photon using the Abelian duality, $F_{\mu\nu}\sim\epsilon_{\mu\nu\rho}\partial^\rho \sigma$. Since the $U(1)$ global symmetry is broken, we expect Goldstone bosons on either side of the duality. For the $\phi$ field, we have massless angular excitations. In this phase the photon remains gapless and it is identified with the Goldstone boson of the $\Phi$ side.

We claim the above results completely generalize to the $n$-node quiver case. Across the duality we have established the fact that the global $U(1)$ symmetry of the $X_{i,i+1}$ bifundamental is identified with the monopole number symmetry of the $i$th node. We begin with the side of the duality where a $U(1)$ global symmetry is unbroken, which corresponds to a positive mass deformation on the $SU$ side and a negative mass deformation on the $U$ side. We will focus on the behavior of a single bifundamental since generalization is straightforward.

When we gap the $Y_{i,i+1}$ bifundamental on the $SU$ side, on the $U$ side this should cause the $X_{i,i+1}$ bifundamental to acquire a vacuum expectation value as a result of the $M_{i,i+1}^2\leftrightarrow-m_{i,i+1}^2$ mass mapping.

Let's take a closer look at the breaking term to account for degrees of freedom. Schematically the interaction term can be written in the form
\begin{align}
V(X_{i,i+1})\sim\Tr\left[\left(X_{i,i+1}\right)_{a_{i}a_{i+1}}\left(X_{i,i+1}^{\dagger}\right)^{b_{i}a_{i+1}}-v^{2}\delta_{a_{i}}^{b_{i}}\right]^2.
\end{align}
With the gauge freedom we can take $\langle X_{i,i+1}\rangle$ to be of the from of \eqref{eq:u vev tw}. This causing the breaking of
\begin{align}
U(K_i)\times U(K_{i+1})\to U(K_i)_{\text{diag}}\times U(k_{i+1}),
\end{align}
corresponding to an overall broken $U(K_i)$ gauge symmetry. Each $X_{i,i+1}$ field has $2K_i K_{i+1}$ total degrees of freedom. Within the $U(K_{i})$ subspace, there are $K_i^2$ flat directions corresponding to ``angular'' excitations, which are consumed by the broken $U(K_i)$ gauge fields to become (two-component) massive ``W-bosons''. The remaining $K_i^2$ scalar degrees of freedom represent ``modulus'' excitations in directions of the potential which are not flat and are thus analogous to Higgs bosons. Additionally, these modes are now adjoint particles since they are charged under $U(K_i)_{\text{diag}}$. This leaves $2K_i k_{i+1}$ degrees of freedom, which acquire a mass through the double trace-like interaction term that is present for Wilson-Fisher scalars. Hence all bifundamental scalar particles are gapped, as they should be.

As with the particle-vortex duality, we would like to show vortices on the $U$ side should be identified with the gapped particles on the $SU$ side of the duality. The gapped $Y_{i,i+1}$ particles are charged under the unbroken $U(1)$ symmetry and carry baryon number. Meanwhile, when the $X_{i,i+1}$ particles acquire a vacuum expectation value, the breaking of the corresponding $U(1)$ subgroup mean vortices associated to that link now have finite mass and are topologically stable since
\footnote{One might worry that we may be able to form other vortex solutions by winding the other broken subsets, say $U(1)_A\subset U(K_{i+1})\times U(K_{i+2})/U(K_i)_{\text{diag}}$. Note however that the interactions force the vacuum expectation value of the associated $U(K_i)$ subgroup of the $X_{i+1,i+2}$ bifundamental to be effectively infinite. This is distinct from $\langle X_{i,i+1}\rangle$  which is presumed to be proportional to the mass deformation and finite. Thus such vortices are significantly heavier than the vortex formed from a winding of the $X_{i,i+1}$ bifundamental and its corresponding gauge groups. Note for the $X_{i,i+1}$ vortex, no matter the mass deformations of bifundamentals to its left, its $U(k_i)$ subgroup will always have a finite vacuum expectation value, and thus the topologically stable vortex solutions can always have finite energy via a winding of this corresponding subgroup.}
\begin{align}
\pi_1\left(U(K_i)\times U(K_{i+1})/U(K_i)_{\text{diag}}\right)\simeq\mathbb{Z}.
\end{align}
Specifically, the vortex configurations correspond to a winding of the broken $U(1)_A\subset U(K_i)\times U(K_{i+1})/U(K_i)_{\text{diag}}$ gauge group as well as the phase of $\langle X_{i,i+1}\rangle$  at spatial infinity. Since the broken subgroup $U(1)_A$ contains the $i$th node's $U(1)$ factor, through the BF term of that node it coupled to the $\tilde{B}_2$ field.\footnote{Since the $X_{i,i+1}$ vortices contain winding under $U(1)_A$ , which is a subgroup of the $U(1)\times U(1)$ gauge symmetries of the $i$th and $(i+1)$th nodes, one might worry that such a vortex also carries flux under the $(i+1)$th gauge group and is thus charged under the $U(1)$ symmetry of the $(i+1)$th node. However, as explained in Appendix \ref{appendix:sym}, the BF coupling is such that nodes to the right of a bifundamental are only coupled via the unbroken gauge group. Hence, although the vortices carry $U(1)$ flux of the $(i+1)$th node, they are only charged under global symmetry of the $i$th node.} This is the same symmetry the gapped $Y_{i,i+1}$ couple to, and thus the two modes should be identified in a manner analogous to what we saw for the particle-vortex duality.

Unlike the particle-vortex duality, the presence of nonzero Chern-Simons terms in the mass deformed phases means variation with respect to dynamical gauge groups imposes a flux attachment condition on the excitations. That is, particles charged under the respective symmetry must be attached to the vortex excitations. This might be modified slightly due to the breaking of the $U$ gauge group, since the broken gauge degrees of freedom will become massive giving an extra term when varying with respect to the corresponding massive gauge degrees of freedom. We leave such analysis for future projects.

\section{Theta Wall Dualities}
\label{sec:thetawalls}

In this section we consider duals to the Chern-Simons theories found on defects in $3+1$-dimensional $SU(N)$ Yang Mills theory when the $\theta$ angle varies as a function of location. Specifically, we look for a dual to \eqref{eq:su_quiver_ym}.

We begin by reviewing a few essential facts about the expected theta dependence in pure $SU(N)$ gauge theories. Such gauge theories are believed to have multiple vacua related to the physics of the theta angle. In each vacuum, physical quantities are not periodic in theta with period $2 \pi$ but instead with period $2 \pi N$. The physical properties of the system are nevertheless $2\pi$ periodic. As we change theta by a single $2 \pi$ period, the true vacuum of the system changes and the physics in the new vacuum at $\theta= \theta_0 + 2 \pi$ is the same as the physics in the original vacuum at $\theta = \theta_0$.

This picture can be most rigourously established at large $N$. In this limit the vacuum energy as a function of $\theta$ is expected to scale as \cite{Witten:1980sp,Witten:1998uka}
\beq \label{single} E(\theta) = N^2 h(\theta/N) \eeq
for some to be determined function $h$. This appears to be inconsistent with the periodicity requirement
\beq E(\theta) = E(\theta + 2 \pi). \eeq
As claimed above, a single vacuum with energy of the form \eqref{single} is expected to be $2 \pi N$ periodic, not $2 \pi$ periodic. This conundrum can easily be solved by postulating that the theory has a family of $N$ vacua labeled by an integer $K$. In this case the vacuum energy in the $K$th vacuum is given by
\beq
\label{kvac}
E_{K}(\theta) = N^2 h((\theta + 2 \pi K)/N).
\eeq
Most of these vacua are meta-stable, the truly stable vacuum for any given $\theta$ is given by minimizing over $K$:
\beq
E(\theta) = N^2 \min_{K} h((\theta + 2 \pi K)/N).
\eeq
The resulting function $E(\theta)$ has the expected $2 \pi$ periodicity. While the energy of (say) the 0-th vacuum keeps increasing as we increase theta from 0 towards $2 \pi$, the energy of the $K=-1$ vacuum at $\theta=2 \pi$ is exactly the same as the energy of the $0$-th vacuum was at $\theta=0$. One expects that a transition from the $0$-th to the $(-1)$-th vacuum is triggered at $\theta=\pi$. While physics in any given vacuum is $2 \pi N$ periodic, the system as a whole, in its true vacuum, is $2 \pi$ periodic.

We are now in a position to discuss the physics of theta interfaces and domain walls. Let us first turn to the case of interfaces. Starting with a confining gauge theory (pure Yang-Mills in this case), one can introduce interfaces across which the theta angle changes by an integer multiple of $2\pi$,
\beq
\Delta \theta = 2 \pi n.
\eeq
The theory is assumed to be everywhere in the true ground state. This means, in particular, that the index labeling the local vacuum state changes by $-n$ units as the theta angle changes by $2 \pi n$. Since the theta angle is a parameter in the Lagrangian, translation invariance is explicitly broken in this theory and we do not expect any Goldstone bosons corresponding to fluctuations of the position of the interface.

As we explained in the introduction, a spatially varying theta gives rise to domain walls on which Chern-Simons theories live. However, anomaly inflow does not constrain the exact Chern-Simons theory. Ref. \cite{Seiberg:2016gmd} has argued that for $|\nabla\theta|\ll \Lambda$ and $|\nabla\theta|\gg \Lambda$, one should expect the TFTs $[SU(N)_{-1}]^n$ and $SU(N)_{-n}$, respectively. Assuming a smooth transition, \eqref{eq:su_quiver_ym} was proposed as a possible CFT to describe the transition between these two extreme cases.

The generic phase of \eqref{eq:su_quiver_ym} is characterized by a partition $\{ n_i \} $ of $n$, that is integers $n_i$ with the property that $\sum_i n_i =n$. Each $n_i$ denotes the number of gauge group factors along the quiver that have been Higgsed down to their diagonal subgroup before we encounter a positive mass squared scalar. For example, $n_1=n$ corresponds to the completely Higgsed $SU(N)_{-n}$ phase associated with the steep wall, $n_i=1$ for $i=1,\ldots,n$ corresponds to the shallow wall with $[SU(N)_{-1}]^n$. The generic phase is given by a TFT based on
\beq
\label{topphase}
\mbox{Phase}\;{ \{ n_i \} }: \quad \quad \prod_{i} SU(N)_{- n_i}.
\eeq
One extra subtlety that arises concerns global symmetries. The scalar fields are bifundamentals under neighboring $SU(N)_{-1}$ gauge group factors. This leaves an overall phase rotation of every single scalar as a global symmetry, for a combined $U(1)^{n-1}$ extra global symmetry from the $n-1$ scalar fields. If these indeed were global symmetries of the parent theory this would lead to unexpected consequences. Most notably, in the fully broken phase the low energy theory on the interface would not just be the topological $SU(N)_{-n}$  Chern-Simons  theory we expect, but would in addition contain $n-1$ massless Goldstone bosons as these extra global symmetries are spontaneously broken in the condensed phase. The proposal of  \cite{Gaiotto:2017tne} is to add extra terms to the action that break these extra global symmetries so that there are no Goldstone bosons. The simplest option to do so is a $\det(Y)$ term for each link\footnote{Of course any power of $\det(Y)$ would do the job in that it is gauge invariant but charged under the global symmetry. For small values of $N$ we need to make use of this freedom. For example, for $N=1$ $\det(Y)=Y$ and we would simply add linear potentials, whereas for $N=2$ we would be adding mass terms. Instead we should add $\det(Y)^4$ and $\det(Y)^2$ respectively in those two cases.}, which is indeed gauge invariant under all $SU(N)_{-1}$ gauge group factors but is charged under overall phase rotations of $Y$. The quiver gauge theories we discussed in the last section do not have these determinant terms added to the potential. The dualities we derive will most naturally apply to the theory without the determinant terms. To connect to the theory of the theta interfaces we will have to add the extra determinant term as a deformation.

In addition to interfaces a second type of co-dimension one defect we can discuss are domain walls. These are already present in a theory with constant theta. They govern the decay of one of the meta-stable vacua of the theory to the true vacuum. In the idealized case, we can consider the theory in a state where we interpolate between two metastable vacua as we move along a single direction, which we once more chose to be the $x_3$ direction. For simplicity we are only interested in configurations which preserve $2+1$ -dimensional Lorentz invariance, that is we focus on flat domain walls. If the theory starts in the $0$-th vacuum as $x_3 \rightarrow \infty$, we can interpolate to the $n$-th vacuum at $x_3 \rightarrow -\infty$. While the state of the system at large negative $x_3$ is not in the true local ground state, this configuration is meta-stable. The false vacuum has to decay via bubble formation, which is governed by the domain wall tension. It has been argued \cite{Witten:1998uka} that the tension of the wall is of order $N$, a fact that is obvious in the holographic realization of these walls which we will turn to in Sec. \ref{sec:thetawall_holo}. At large $N$ this means that decay of the meta-stable vacuum is $e^{-N}$ suppressed. In addition the difference in vacuum energies will exert a pressure on the domain wall generically causing the wall to move. But since the pressure difference is order $N^0$, whereas the domain wall tension is order $N$, the domain wall can be treated as static in the large $N$ limit.

As far as the anomalies are concerned, the analysis of \cite{Gaiotto:2017tne} generalizes to the case of walls: the gauge theory on the defect should be the same whether we are forced to jump $n$ vacua because of a $2 \pi n$ jump in $\theta$, or whether we study a dynamical wall that interpolates between two $n$-separated vacua in a theory with fixed $\theta$. The main difference appears to be that this time the wall is dynamical with a finite tension. Most notably, this implies that we should have (at large $N$) a massless scalar living on the wall whose expectation value gives the location of the wall. It being the Goldstone boson of broken translations, the scalar has an exact shift symmetry that protects it from becoming massive. While interfaces were characterized by a free function $\theta(x_3)$ and were only loosely characterized into shallow and steep, for walls it is much easier to characterize the moduli space of allowed configurations. We have a total of $n$ discrete jumps from one vacuum to the next. When the walls are widely separated, we should have $n$ separate walls connecting two neighboring vacua each. In this limit, we should have a total of $n$ translational modes as the different basic walls can presumably move independently. The gauge theory living on these widely separated walls should be $[SU(N)_{-1}]^n$ as above together with these decoupled light translational modes. This is indeed what follows from the analysis of Acharya and Vafa in the closely related case of ${\cal N}=1$ supersymmetric gauge theories \cite{Acharya:2001dz} (see also \cite{Dierigl:2014xta}). The other extreme is when all $n$ walls coincide and we have a single wall across which we jump by $n$ vacua, presumably governed by a single $SU(N)_{-n}$ gauge theory and a single translational mode.

To summarize, note that we are still characterizing the phases by partitions $\{ n_i \}$ of $n$ and the gauge theory on the wall is once again governed by the topological field theory \eqref{topphase}. In addition we have the  decoupled translational modes. At finite $N$ the walls no longer correspond to static configurations as they will be pushed around by the pressure differences, making them generically much harder to study than the case of interfaces. The reason we discuss them at all as that, at large $N$, they have a very simple holographic realization which we will employ in what follows to check our dualities.

\subsection{Theta Wall Dualities via 3d Bosonization}
\label{sec:thetawall_3d}

We now consider possible duals to \eqref{eq:su_quiver_ym} via 3d bosonization. Fortunately, such a theory can easily be constructed from the non-Abelian linear quiver dualities.\footnote{As touched upon earlier, if one tried to derive such a quiver using only Aharony's dualities, one would inevitably run into violations of the flavor bound. Thus it appears the interactions between links which come from the master duality are a necessity.} Consider the $n$ node linear quiver, \eqref{eq:general_n}, and take
\begin{subequations}
\label{eq:thetawall_values}
\begin{align}
1=k_{1} & =k_{2}=\cdots=k_{n}\\
N=N_{1} & =N_{2}=\cdots=N_{n}.
\end{align}
\end{subequations}
This satisfies all flavor bounds of the derivation given above since $k_{i}\geq0$ and $N=N_{j}\geq N_{j+1}=N$. In this case the dual quiver theories become
\begin{subequations}
\begin{align}
\text{Theory A:}\qquad & \left[SU(N)_{-1}\right]^{n}\times\prod_{p=1}^{n-1}\text{bifundamental }Y_{p,p+1}\\
\text{Theory B:}\qquad & \prod_{p=1}^{n-1}\left[U(p)_0\times\text{bifundamental }X_{p,p+1}\right]\times U\left(n\right)_{N}
\end{align}
\end{subequations}
and the relevant mass deformations for all bifundamentals taken positive/negative are given by
\begin{subequations}
\begin{align}
(\text{A1})\qquad M_{i,i+1}^{2}>0:\qquad & \left[SU(N)_{-1}\right]^{n}\\
(\text{A2})\qquad M_{i,i+1}^{2}<0:\qquad & SU(N)_{-n}\times\prod_{p=1}^{n-1}\left[SU(0)_{-p}\right]\\
(\text{B1})\qquad m_{i,i+1}^{2}<0:\qquad & \left[U(1)_{N}\right]^{n}\\
(\text{B2})\qquad m_{i,i+1}^{2}>0:\qquad & U(n)_N\times\prod_{p=1}^{n-1}\left[U(p)_0\right].
\end{align}
\end{subequations}
The topological sector of Theory A matches the TFTs we set to find at the outset, $SU(N)_{-n}$ and $[SU(N)_{-1}]^n$. In addition both sides have decoupled massless modes that also match. We assume that the non-Abelian part of $U(p)_0$ confines at low energies and is therefore gapped. This implies that the dynamics of the confining gauge group should have no effect on the physics at scales well below the gap. The $U(1)$ part however gives rise to a light photon for every level $0$ unitary gauge group. On the $SU$ side these light photons map to Goldstone bosons. In a theory with $N_s < N$ scalars charged under a $SU(N)$ gauge symmetry a full global $U(N_s)$ flavor symmetry is unbroken as the gauge group is broken to $SU(N-N_s)$ by a scalar vacuum expectation value. The broken gauge generators can be used to compensate any flavor rotation. In the special case of $N_s=N$, which is of interest to us here, the $U(1)$ part of the flavor symmetry however is broken and so we will get a corresponding Goldstone boson. In order to keep track of these light scalars we denote the Goldstone bosons as $SU(0)_{-p}$ theories, which continue to be ``level-rank'' dual to the $U(p)_0$ factors of Theory $(\text{B2})$; either theory denotes a decoupled light scalar mode. Including these factors we see that there is a perfect matching both between the topological sector and the decoupled light modes.

One should note that these extra massless Goldstone bosons are exactly the ones that in the theory of theta interfaces have been eliminated by the $\det(Y)$ potentials. As it stands, our quiver duality applies to \eqref{eq:su_quiver_ym} without these extra determinant terms. Since the global $U(1)$ baryon number symmetries under which $\det(Y)$ is charged map to monopole symmetries on the $U$ side, the corresponding dual operator is a monopole operator. Adding this monopole operator to the theory should lead to confinement of the $U(1)_0$ factors together with their non-Abelian counterparts and hence remove the massless photons associated to these factors from the spectrum, just as we removed their dual Goldstone bosons on the $SU$ side.

Given the fact that most of the gauge group factors on the $U$ side confine, we can further simplify the low energy description of this side of the duality. The confining groups cause the bifundamental matter and antimatter to form ``mesons''.  If the matter/antimatter is still charged under some gauge group with nonzero Chern-Simons level, the meson transforms as an adjoint under said gauge group as conjectured in \eqref{eq:u_quiver_ym}.
For phase $(\text{B2})$, there are the adjoints formed from $X_{n-1,n}^{\dagger}X_{n-1,n}$ since the $\left(n-1\right)$th node confines. It is difficult to say if bound states such as $X_{n-1,n}^{\dagger}X_{n-2,n}^{\dagger}X_{n-2,n}X_{n-1,n}$, which are also adjoints under the $U(n)_N$ gauge group, would be stable or if it would split into separate particles $X_{n-2,n}^{\dagger}X_{n-2,n}$ and $X_{n-1,n}^{\dagger}X_{n-1,n}$. If we assume the latter, there is only a single light adjoint scalar charged under the $U(n)$ considered above. \footnote{Note that when the bifundamental scalars on the $SU$ side acquire a negative vacuum expectation value, by assumption this breaks their gauge symmetry down to the common diagonal symmetry group and causes the Higgsed bifundamentals to become adjoint particles. Thus we get gapped adjoint particles on both sides of the duality for phase 2 considered above.} We also would want to conjecture that there are no additional neutral mesons that become light together with the adjoint; such extra light matter is not accounted for on the $SU$ side of the duality. With these dynamical assumptions our quiver duality boils down to the one we advertised in the introduction
\begin{align}
\label{eq:u_quiver_ym_reloaded}
[SU(N)_{-1}]^n + \text{bifundamental scalars} \qquad\leftrightarrow\qquad U(n)_{N} + \text{adjoint scalars}
\end{align}
with a $\det(Y)$ potential for all the bifundamental scalars on the $SU$ side implied.

Unlike the quiver dualities, which we derived from gauging global symmetries, the duality \eqref{eq:u_quiver_ym} only follows upon making extra dynamical assumptions regarding the confining mechanism. We can give extra evidence for this duality by, once again, looking at the phase structure. On the $U$ side the various massive phases are realized by adding mass squared terms that give expectation values to the adjoint scalar (or remove it completely) together with $\Tr X^k$ terms in the potential. We can always chose a gauge in which the scalar expectation value is diagonal, so the generic expectation values is characterized by the $n$ eigenvalues of the scalar expectation value. Due to the presence of the interaction terms, it is possible to have none of the eigenvalues coincide, in which case the gauge group is $[U(1)_N]^n$. But whenever two or more eigenvalues coincide, we do get an enhanced unbroken subgroup. Once again the most general phase is encoded in partitions $\{ n_i \}$ of $n$, where each integer $n_i$ denotes the multiplicity of a given eigenvalue. The generic phase is given by $n_i=1$ for all $i$, whereas the case of $n$ coincident eigenvalues with a single $U(n)_N$ gauge group factor corresponds to $n_1=n$. The generic partition corresponds to
\beq
\label{bulktopphase}
\mbox{Phase}\;{ \{ n_i \} }: \quad \quad \prod_{i} U(n_i)_{N}.
\eeq
Reassuringly, this is exactly the level-rank dual gauge group of what we found for the quiver theory, \eqref{topphase}. In the next section we will give further support for the validity of this duality, at least in the large $N$ limit, using holography.

\subsection{Theta Wall Dualities via Holography}
\label{sec:thetawall_holo}
\label{sec:jensen}

Now we turn to the holographic proof of the duality. Our work will follow closely the stringy embedding of bosonization presented in \cite{Jensen:2017xbs} based on the earlier string theory realization of level-rank duality in \cite{Fujita:2009kw}. In this construction the holographic duality between field theory and supergravity becomes, at low energies, the purely field theoretic bosonization duality.

One starts with a well known holographic pair. The work of \cite{Fujita:2009kw,Jensen:2017xbs} employs the original holographic duality \cite{Maldacena:1997re} between ${\cal N}=4$ super-Yang Mills (SYM) and type IIB string theory on AdS$_5$ $\times$ $S^5$. We then deform the theory in such a way that all, or at least most, degrees of freedom gap out and one is left with a non-trivial topological field theory (in the case of level-rank) or conformal field theory (in the case of bosonization) in the infrared. Following the same deformations in the dual gravity solution one finds that the spectrum of most supergravity excitations also gets gapped out. The only remaining low energy excitations are localized on a probe brane. These probe degrees of freedom in the bulk are found to be related to the boundary degrees of freedom by the desired field theory duality.

\subsubsection*{Review of Holography Applied to 3d Bosonization}

Let us first briefly review the case of level-rank. Starting with ${\cal N}=4$ SYM one can go to $2+1$ dimensions via compactifying the theory on a circle of radius $R$. With anti-periodic boundary conditions for the fermions in the theory, all fermionic Kaluza Klein modes pick up masses of order $1/R$ and the scalars then pick up masses of the same order via loop corrections. At energies below $1/R$ we are left with pure Yang-Mills in $2+1$ dimensions, which is believed to confine. The theory is gapped with gap of order $1/R$. This is not quite yet the theory we want, the IR is trivial rather than a non-trivial Chern-Simons TFT.

To produce the desired Chern-Simons terms we need to introduce the theta angle. Like all coupling constants in the Lagrangian, the theta angle in $3+1$ dimensional gauge theories is usually introduced as a position independent constant, but it can be promoted to a non-trivial background field. What we need here is a theta angle that linearly changes by $2 \pi n$ as we walk around the circle once. Since theta is only well defined modulo $ 2 \pi$ this is consistent as long as $n$ is an integer. The $\theta F \wedge F$ term in the Lagrangian with constant theta gradient can be integrated by parts to turn into a $2+1$ dimensional  Chern-Simons  term with level $-n$. So in short, ${\cal N}=4$ SYM with anti-periodic boundary conditions for fermions and a constant theta gradient gives rise to a gapped $2+1$ dimensional theory which, at low energies, is well described by an $SU(N)_{-n}$ Chern-Simons theory.

These deformations are easily repeated in the holographic dual. The compactification with anti-periodic boundary conditions for fermions is dual to the cigar geometry of \cite{Witten:1998zw}, that is a doubly-Wick rotated planar Schwarzschild black-hole where the compact time direction of the Euclidean black hole plays the role of the compact spatial directions, whereas one of the directions along the planar ``horizon" becomes the new time direction. Most importantly, the radial coordinate in this cigar geometry truncates at a finite value $r=r_*$ where the compact circle contracts. Consequently this geometry acts as a finite box and so indeed all supergravity fluctuations exhibit a gapped spectrum \cite{Witten:1998zw} with a mass gap of order $1/R$. In order to retain a non-trivial topological sector we still need to implement the spatially varying theta angle. The theta angle is set by the near boundary behavior of the bulk axion field, so we are looking for a supergravity solution where the axion asymptotes to $a \sim n y/R$. Here $y$ denotes the coordinate along the circle direction and $a=n y/R$ is an exact solution to the axion equation of motion in the cigar background. As long as we are only interested in the $n \ll N$ limit we can ignore the backreaction of the axion on the background geometry and $a=ny/R$ appears to be the full solution to the problem. The only remaining issue is that the axion field strength $f_y = \partial_y a = n/R$ in the bulk has to be supported by a source. This source can be introduced by locating $n$ D7 branes, wrapping the entire internal $S^5$, at the tip of the cigar at $r=r_*$. This stack of D7 brane introduces new degrees of freedom in the bulk. The scalar fields corresponding to fluctuations of the D7 away from the tip are massive due to the geometry of the cigar. Like all other geometric fluctuations they have mass of order $1/R$. The only other degree of freedom introduced by the $n$ D7 branes is the worldvolume gauge field. The latter acquires a Chern-Simons term of level $N$ from the Wess-Zumino coupling to the $N$ units of background 5-form flux through the $S^5$. Lo and behold, the low energy description of the holographic bulk dual is simply a $U(n)_N$  Chern-Simons  gauge theory living on the D7 branes. Comparing low energy descriptions on both sides, AdS/CFT boiled down to level-rank duality for the emerging TFTs.

The last step in order to derive 3d bosonization rather than level-rank from this construction is to add extra light matter into the theory. This can be easily accomplished using flavor probe branes \cite{Karch:2002sh}. In the construction put forward in \cite{Jensen:2017xbs} an extra probe D5 adds fermionic matter localized on $2+1$ dimensional defects in the $3+1$ dimensional theory. These defects live at points in the circle direction, so at low energies they simply become light fermions coupled to the $SU(N)_{-n}$  Chern-Simons  gauge fields. The same probe branes can be argued, from the bulk point of view, to add scalar matter to the dual $U(n)_N$  Chern-Simons  gauge theory. Instead of simply giving us level-rank, in this case holography, at low energies, reduces to the basic non-Abelian 3d bosonization duality.

\subsubsection*{Holographic Realization of Theta Walls}

To holographically realize the field theory theta domain walls we just reviewed we need to start with a holographic duality for a confining $3+1$ dimensional theory and then simply once again follow the field theory deformation corresponding to turning on theta in the bulk. The simplest realization of a confining $3+1$ gauge theory with a gravity dual is Witten's black hole \cite{Witten:1998zw}. This is almost the same construction we employed previously, but lifted one dimension up. We start with a 5d gauge theory, maximally supersymmetric YM with gauge group $SU(N)$, and compactify it on a circle with anti-periodic boundary conditions. The dual geometry has once again the basic shape of a cigar, and the explicit supergravity solution is given by
\beq \nonumber
ds^2 = \left ( \frac{u}{L} \right )^{3/2} \left( \eta_{\mu \nu} dx^{\mu} dx^{\nu} + f(u) d y^2 \right)
+ \left ( \frac{L}{u} \right )^{3/2} \left ( \frac{du^2}{f(u)} + u^2 d \Omega_4^2 \right ),
\eeq
\beq
\label{cigar}
e^{\phi} = g_s \left ( \frac{u}{L} \right )^{3/4}, \quad F_4 = dC_3 = \frac{2 \pi N}{V_4} \epsilon_4,
\quad f(u) = 1 - \frac{u_*^3}{u^3}.
\eeq
Here $x^{\mu}$ are the 4 coordinates of $3+1$ dimensional Minkowski space, $y$ is the circle direction we compactified to go from $4+1$ to $3+1$ dimensions. $\phi$ is the dilaton field, $F_4$ the RR 4-form field strength. $\Omega_4$ is the internal 4-sphere, with $d\Omega_4^2$, $\epsilon_4$ and $V_4 = 8 \pi^2/3$ its line element, volume form and volume respectively. The string coupling $g_s$ and the string length $l_s$ are the parameters of the underlying type IIA super-string theory. $L$ sets the curvature radius of the solution, it is determined by Einstein's equations to be $L^3 = \pi g_s N l_s^3$. Last but not least $u_*$ is the location of the tip of the cigar, it is related to the periodicity $2 \pi R$ of the compactification circle by $R = \frac{2}{3} L^{3/2} u_*^{-3/2}$.

The holographic realization of turning on a constant theta angle has been worked out in \cite{Witten:1998uka}. The theta angle is dual to the Wilson line of the bulk RR 1-form $C_{\mu}$ along the compact $y$ direction:
\beq
\label{thetawilson}
\int_{S^1} C = \theta +\ldots
\eeq
where the ellipses denote terms with negative powers of $u$, that is terms that vanish near the boundary.
The Wilson line is gauge invariant modulo $2 \pi \mathbb{Z}$, so theta is indeed an angle.
Using Stokes's law, we can rewrite the condition \eqref{thetawilson} as
\beq
\label{thetaf}
\int_D F = \theta + 2 \pi K.
\eeq
Here $F=dC$ is the field strength associated with the RR one-form and $D$ is the cigar geometry, which has the topology of a disc. Since $\int_D F$ is a well-defined real number whereas $\theta$ is an angle, we have a $2 \pi K$ ambiguity in $F$ where $K$ is an integer. For a given theta there is more than one bulk solution for $F$, characterized by $K$. This is responsible for the multi-branched structure of the allowed ground states which we expect to find. Physics in any given one of the branches is only periodic in $2 \pi N$, the actual periodicity of $\theta$ is $2 \pi$ as it should be. We simply jump to a different branch.

For generic theta it is non-trivial to solve the supergravity solutions subject to the constraint \eqref{thetaf}. But a very simple solution can once more be found \cite{Witten:1998uka,Bartolini:2016dbk} in the probe limit $(\theta + 2 \pi K) \ll N$, or in other words $K/N \ll 1$. In this limit one can neglect the backreaction of the axion on the background geometry. Newton's constant is of order $1/N^2$ in units where the curvature scale $L=1$, whereas the axion action and hence its stress tensor is of order 1 in the large $N$ counting. The only non-trivial equation left to solve is Maxwell's equation for $C_1$ in the background geometry \eqref{cigar} subject to the boundary condition \eqref{thetaf}. The solution is
\beq
\label{csol}
C_1 = \frac{f(u)}{2 \pi R} (\theta + 2 \pi K) dy.
\eeq
The integer $K$ is the bulk manifestation of the $K$-th vacuum. In fact, plugging the solution \eqref{csol} back into the action we find that the vacuum energy density of the $K$-th vacuum has exactly the expected form from \eqref{kvac} with \cite{Bartolini:2016dbk}
\beq
h(\theta/N) = - \frac{2 N^2 \lambda}{3^7 \pi^2 R^4}
\left [ 1 - 3 \left ( \frac{\lambda}{4 \pi^2}
\right )^2 \left ( \frac{\theta + 2 \pi K}{N} \right )^2 \right ]
\eeq
where $\lambda = g_{YM}^2 N = 2 \pi g_s l_s N/R$ is the 't Hooft coupling.

\begin{figure}
\begin{centering}
\includegraphics[scale=0.4]{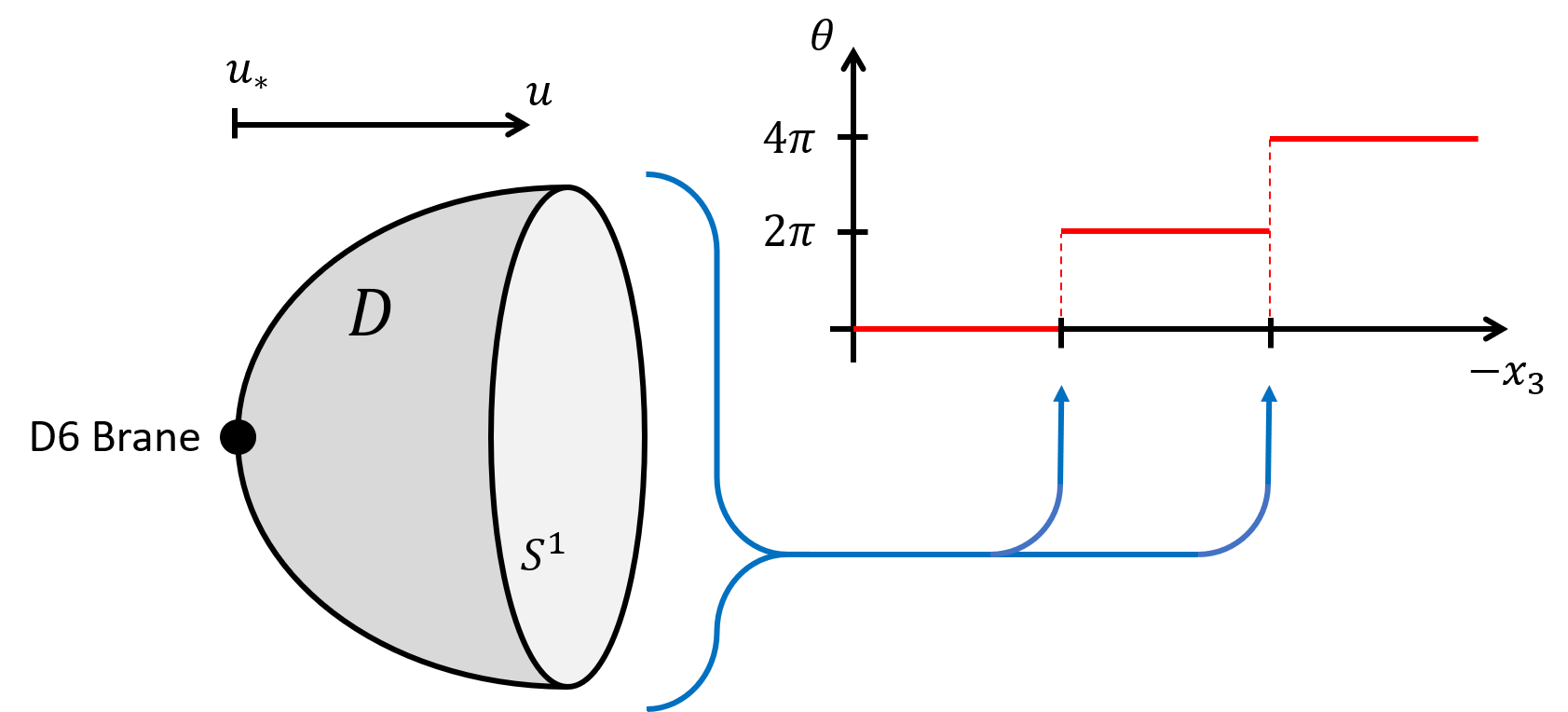}
\par\end{centering}
\caption{Configuration of D6 branes. \label{fig:d6_Branes}}
\end{figure}

While it is not obvious to us how to realize interfaces in this setup, the holographic dual for a domain wall has already been proposed in \cite{Witten:1998uka}. A jump in vacuum, according to \eqref{thetaf}, requires a jump in $\int_D F$, which in turn requires a source magnetically charged under the RR 1-form. The naturally stringy object carrying the appropriate RR charge is a D6 brane. The D6 brane needs to wrap the entire internal $S^4$ as well as the 3d Minkowski space spanned by $t$, $x_1$ and $x_2$. It is localized in the $x_3$ direction as well as on the cigar geometry $D$. From the induced metric of a D6 sitting at a fixed position $u$ and wrapping $M^{2,1} \times S^4$ is we can infer that  the D6 Lagrangian density $e^{- \phi} \sqrt{-g_I}$ reads
\beq
{\cal L} \propto u^{5/2}
\eeq
meaning that the D6 brane experience a potential pulling it to smaller values of $u$: the D6 brane will sink to the tip of the cigar, see Fig. \ref{fig:d6_Branes}.

Let us first discuss the case of a single D6 brane. Without loss of generality, we can place the D6 at $x_3=0$. If we denote by $D_-$ the cigar/disc spanned by $(y,u)$ at a fixed negative $x_3$ and $D_+$ the cigar/disc at a fixed positive $x_3$, then the analog of the magnetic Gauss' law for the D6 brane reads
\beq
\int_{D_+} F - \int_{D_-} F = 2 \pi.
\eeq
Comparing with \eqref{thetaf} we see that this means that $(\theta + 2 \pi K)$ jumps by $2 \pi$ as we cross, in the field theory, the bulk $x_3$ location of the D6-brane. This also implies that the vacuum energy of the theory jumps across the D6. Furthermore, the D6 brane is clearly dynamical. The $x_3$ position of the D6 brane is a dynamical field. Since the metric is independent of $x_3$ the corresponding worldvolume scalar is massless. These facts together clearly identify the D6 brane as the domain wall between the $j$-th and $(j+1)$-th vacuum \cite{Witten:1998uka} at a fixed theta angle. The general wall in which we jump from the $j$-th vacuum to the $(j+K)$-th simply corresponds to $K$ coincident D6 branes. We can pull apart the stack of $K$ D6 branes to obtain a configuration of walls where the vacuum jumps one unit at a time at well-separated locations in the $x_3$ direction.

It is fairly straightforward to determine the low energy physics in the bulk. The background geometry once more truncates at a finite radial position, $u=u_*$. Correspondingly all supergravity modes are gapped. The only degrees of freedom surviving are the ones localized on the D6 branes. For a stack of $n$ coincident D6 branes, these worldvolume degrees of freedom are a $U(n)$ gauge field as well as 3 adjoint scalars corresponding to motion of the stack into the $u$, $y$ and $x_3$ direction. The $u$ and $y$ fluctuations are massive due to the cigar geometry just as we reviewed above in section \ref{sec:jensen}. The $x_3$ scalar, however, is massless. The worldvolume gauge field picks up a Chern-Simons term of level $N$ from the Wess-Zumino coupling of the worldvolume gauge field to the $N$ units of 4-form flux. So the low energy dynamics in the bulk is governed by a $U(n)_N$ gauge theory with a single massless adjoint representation scalar. Holographic duality implies that this is an equivalent representation of the quiver gauge theory with the additional $n-1$ translational modes associated with the domain walls at least in the large $N$ limit.

Note that this way we almost landed on the duality \eqref{eq:u_quiver_ym}. There is however a small difference. On the quiver side, we have the extra light modes corresponding to the translational motion of the domain walls. As we argued before, we expect these to be present at large $N$. At any finite $N$ the domain walls would no longer be static. The phases of the quiver theory are still given by \eqref{topphase} as long as one accounts for the extra decoupled translational modes. The analogous statement on the $U$ side of the duality is that the adjoint scalar governing the position of the stack of probe branes this time corresponds to a flat direction. The various phases are still parametrized by the $n$ eigenvalues of the scalar matrix $\langle x_3 \rangle$. But this time instead of having to add deformations to the potential we have a moduli space of vacua where we can freely dial the expectation values of $x_3$. The eigenvalues of $\langle x_3 \rangle$ simply correspond to D6 positions and the enhanced gauge symmetries we encounter for coincident eigenvalues simply arise from coincident D6 branes. The gauge groups of the various phases once again are given by \eqref{bulktopphase}, but since the scalar potential was exactly flat this time each gauge group factor comes with an extra massless adjoint. In the generic case where the gauge group is $[U(1)_N ]^n$ these extra massless adjoints map exactly to the $n$ translational modes we identified on the quiver side. Surely the exactly flat potential for the probe scalar is also an artifact of the large $N$ limit and any bulk quantum corrections would lift this flat direction. Furthermore, the phase where the adjoint gets a positive mass squared is not easily realized in the brane picture. Modulo these extra light scalars, matching on both sides, the holographic construction exactly reproduces our conjectured duality
\eqref{eq:u_quiver_ym}.

\section{Discussion and Conclusion}
\label{sec:conclusion}

In this work, we have developed the methodology for dualizing linear quiver gauge theories with bifundamental scalars and argued that they can be viewed as the non-Abelian generalization of particle/vortex duality. Crucial to this is the interaction terms, which couple scalars living on adjacent links and propagate the symmetry breaking pattern down the quiver in a unidirectional manner. This is required to ensure the mass deformed phases are level-rank dual to each other.

We then specialize this general framework to the study of domain walls that arise in $3+1$ dimensional Yang-Mills theory with a spatially varying theta angle. In addition, we embed this special case in string theory and study the duality holographically. We find a novel duality between a theory with bifundamental matter and one with adjoint matter, schematically given by  \eqref{eq:u_quiver_ym}. Let us comment on the similarities and differences between these two approaches.

From the setup in ref. \cite{Gaiotto:2017tne}, we expect the bifundamental scalars on the $SU$ side of the duality to interpolate between a smoothly varying and a sharp domain wall/interface. However, the pure field theoretic quiver approach of Sec. \ref{sec:building_quivers} makes opaque the geometric interpretation of a physical wall located in space. The complementary geometric approach of Sec. \ref{sec:jensen} makes this manifest: Higgsing a bifundamental is literally removing a D6 brane (i.e. domain wall) from a stack and moving it to a different physical location in space. Widely separated D6s correspond to the smoothly varying phase, reinforcing our intuition of the theory at small $| \nabla \theta |$.


When the walls on top of one another in the holographic duality we have new light matter one both sides of the duality, but it manifests itself in a very different manner. On the $U$ side the extra matter enhances the gauge symmetry. Meanwhile, on the $SU$ side the bifundamentals just become additional massless scalars. This may seem peculiar given the fact the bifundamental degrees of freedom match quite nicely when the walls are separated (albeit by construction). But this is precisely the behavior we would expect in our 3d bosonization duality. As we learned from the particle-vortex duality generalization, it is not actually the bifundamental degrees of freedom which should match on either side of the duality but rather particles and vortices. The very same mismatch of particle degrees of freedom is present in the bosonic particle-vortex duality as well.

In fact, the matching of the particle and vortex degrees of freedom is very nicely realized in the holographic duality. The dual of the baryons in the bulk are based on the standard holographic construction of the baryon vertex \cite{Witten:1998xy}, very similar to what was found in \cite{Jensen:2017xbs}. Namely, they are D4 branes wrapping the $S^4$ and also extended along the time direction. In order to be neutral, fundamental strings run from the D4 branes to the D6 branes on which they live. Furthermore, the D6 branes dissolve the D4 branes turning them into magnetic flux (it is more energetically favorable and has the same quantum numbers). The attachment of $N$ fundamental strings is analogous to particle/flux attachment. It can be argued that the $N$ fundamental strings cannot end on the same D6 brane. Hence when the D6 branes get separated, the monopoles must also pick up a mass since the fundamental strings must stretch from one D6 brane to another -- providing more evidence that lines of flux can end on domain walls. This is also in nice agreement with the behavior which occurs on the $SU$ side where the bifundamentals are interpreted as strings which stretch from one brane to another and thus both acquire a mass proportional to the separation between branes.

One may wonder whether our quiver dualities can be useful in the context of deconstruction, following the recent work of \cite{Aitken:2018joz}. There it was shown that Abelian quiver dualities can be lifted to dualities in 3+1 dimensions. It would be very interesting to do this in the non-Abelian case. One important ingredient in this construction is the use of ``all scale" versions of the duality, following the construction of \cite{Kapustin:1999ha} in the supersymmetric case. We'd like to point out that at least for two-node quivers, our method of gauging global flavor symmetries does allow us to give all scale versions of the non-Abelian duality. Say we want a dual for $SU(N)_{k}$ with $N_f$ fermionic flavors with a finite gauge coupling. Since the gauge coupling is dimensionful we are describing a theory with a non-trivial RG running. It interpolates between a free theory in the UV and a strongly coupled CFT in the IR. We can obtain this theory by starting with $N N_f$ free fermions and gauging a $SU(N)$ subgroup of the global $SU(NN_f)$ flavor symmetry. At this stage we can add both the Chern-Simons as well as the Maxwell kinetic terms. The original theory of $NN_f$ free fermions has dual descriptions in terms of a $U(K)_1$ gauge theory coupled to $N N_f$ scalars. Modulo flavor bounds $K$ is a free parameter. The global flavor symmetry simply rotates the scalar flavors in this dual. Promoting a $SU(N)$ subgroup to be dynamical we end up with a $U(K)_1$ $\times$ $SU(N)_{k}$ gauge theory with $N_f$ bi-fundamentals. While the $U(K)$ factor has infinite coupling, the $SU(N)_k$ factor has a finite Maxwell term which maps directly to the Maxwell term of the same $SU(N)_k$ factor on the dual side. This way we did construct a non-Abelian all scale dual to $SU(N)_k$ with fermions. Unfortunately it is not yet clear how to generalize this construction to more interesting quivers.

\section*{Acknowledgments}

We would like to thank Aleksey Cherman, Kristan Jensen, Zohar Komargodski, Brandon Robinson, and David Tong for useful discussions. This work was supported, in part, by the U.S.~Department of Energy under Grant No.~DE-SC0011637.


\appendix

\section{Building Non-Abelian Linear Quivers}
\label{appendix:dual}

Here we provide further details of our construction of the non-Abelian linear quivers.

\subsection{Other Forms of the Duality}
\label{appendix:duals}

The second of Aharony's dualities is given by taking the $N_s=0$ of the master duality, \eqref{eq:master_dual}, which gives
\begin{subequations}
\begin{align}
\mathcal{L}_{SU} & = i\bar{\psi}\Dslash_{b^\prime+C+\tilde{A}_1}\psi-i\left[\frac{N_f-k}{4\pi}\text{Tr}_{N}\left(b^{\prime}db^{\prime}-i\frac{2}{3}b^{\prime3}\right)+\frac{N}{4\pi}\text{Tr}_{N_f}\left(CdC-i\frac{2}{3}C^{3}\right)\right]\nonumber \\
 & -i\left[\frac{N\left(N_f-k\right)}{4\pi}\tilde{A}_{1}d\tilde{A}_{1}\right],\\
\mathcal{L}_{U} & =\left|D_{c+C}\phi\right|^{2}-i\left[\frac{N}{4\pi}\text{Tr}_{k}\left(cdc-i\frac{2}{3}c^{3}\right)-\frac{N}{2\pi}\text{Tr}_{k}\left(c\right)d\tilde{A}_{1}\right].
\end{align}
\end{subequations}
Performing the using $\tilde{c}\to\tilde{c}+\tilde{A}_1$ shift, canceling common factors on either side of the duality, and defining $E_\mu\equiv C_\mu + \tilde{A}_{1\mu}$, we end up with
\begin{subequations}
\begin{align}
\mathcal{L}_{SU} & =i\bar{\psi}\Dslash_{b^{\prime}+E}\psi-i\left[\frac{N_f-k}{4\pi}\text{Tr}_{N}\left(b^{\prime}db^{\prime}-i\frac{2}{3}b^{\prime3}\right)+\frac{N}{4\pi}\text{Tr}_{N_f}\left(EdE-i\frac{2}{3}E^{3}\right)\right]\\
\mathcal{L}_{U} & =\left|D_{c+E}\phi\right|^{2}-i\left[\frac{N}{4\pi}\text{Tr}_{k}\left(cdc-i\frac{2}{3}c^{3}\right)\right].
\end{align}
\end{subequations}
This yields the duality \eqref{eq:aharony dual tw} which we use in going from theory D to theory E. Note that in promoting $\tilde{A}_1$ into a dynamical field, we again introduce a new global symmetry. We will call the background gauge field associated with said symmetry $\tilde{B}_1$.

Returning to the master duality, in the main text we could have combined the $U(1)$ and $SU(N_f)$ global symmetries into the definition of $E_{\mu}=C_{\mu}+\tilde{A}_{1\mu}\mathds{1}_{N_f}$. For the purposes of deriving the non-Abelian linear quivers, this was not necessary. For completeness, we show the form of the duality here, because it gives the explicit master duality with all background fields in its most succinct form. It is also convenient to define the $U(N_s)$ gauge field $H_\mu\equiv B_\mu+\tilde{A}_{1\mu}+\tilde{A}_{2\mu}$. This leaves us with a duality of the form
\begin{subequations}
\begin{align}
\mathcal{L}_{SU} & =\left|D_{b^{\prime}+H}\phi\right|^{2}+i\bar{\psi}\Dslash_{b^{\prime}+ E }\psi+\mathcal{L}_{\text{int}}-i\left[\frac{N_f-k}{4\pi}\text{Tr}_{N}\left(b^{\prime}db^{\prime}-i\frac{2}{3}b^{\prime3}\right)\right]\nonumber \\
 & -i\left[\frac{N}{4\pi}\text{Tr}_{N_f}\left(EdE-i\frac{2}{3}E^{3}\right)\right],\\
\mathcal{L}_{U} &= \left|D_{c+E}\Phi\right|^{2}+i\bar{\Psi}\Dslash_{c+H}\Psi+\mathcal{L}_{\text{int}}^{\prime}-i\left[\frac{N}{4\pi}\text{Tr}_{k}\left(cdc-i\frac{2}{3}c^{3}\right)\right].
\end{align}
\end{subequations}
This makes the $U(N_s)\times U(N_f)$ global symmetry explicit.

\subsection{Global Symmetries}
\label{appendix:sym}

Here we discuss the matching of the global symmetries across the dualities in more detail.

In our four node example of Sec. \ref{sec:building_quivers}, when stepping from theory A to B and subsequently from theory B to C, there is an implicit matching that occurs between the two equivalent Lagrangians we call theory B. To follow the global symmetries all the way through from theory A to E, it is necessary to look at this matching more carefully. This implicit matching also occurs for every intermediate theory as well (i.e. theories C and D for the four node case). We will call these two equivalent descriptions theory B' and B" and use a similar notation to describing the matching of C and D as well.

To start, let us explicitly consider the matching of theory B. Specifically, theory B' is what we get from using \eqref{eq:aharony dual 4} with theory A. Meanwhile, theory B" is what we would like to apply the master duality in the form of \eqref{eq:master dual tw} to get out what we now call theory C'. When matching these theories to one another, what was the color gauge symmetry of theory B' gets mapped to the (promoted) $U(k_3)$ flavor symmetry of theory B". Meanwhile, the flavor symmetry of theory B' becomes the color symmetry of theory B".

It will be helpful to consider in closer detail how we are matching all the gauge fields to which the fermions couple. The fermion couplings for the two theories are given by
\begin{align}
(\text{B'}):\qquad & i\bar{\Psi}\Dslash_{c+B+\tilde{A}_1}\Psi\\
(\text{B"}):\qquad & i\bar{\psi}\Dslash_{b^\prime+C+\tilde{A}_1-\tilde{A}_2}\psi.
\end{align}
With a slight abuse of notation, the dynamical/background gauge fields of theory B' and B" denoted above are completely distinct and must be matched. The matching of the gauge fields associated with gauge and global symmetries are shown in Table \ref{tab:matching_theory_B}. Note the $\tilde{A}_1$ field belonging to $\Psi$ is matched to $\tilde{B}_2$ field of the subsequent node.

This careful matching allows us to focus on how the global $U(1)$ symmetry gets transferred through the dualities. In the duality relating theory A and B', the $U(1)$ global symmetry is associated with $\tilde{A}_1$. For theory A this shows up as a baryon-number symmetry and in theory B' it appears as monopole-like symmetry which couples to the $U$ field flux. From Table \ref{tab:matching_theory_B}, we identify this symmetry with the $U(1)$ monopole symmetry of theory B", where the associated background gauge field is $\tilde{B}_2$. Recall, this is the new global $U(1)$ symmetry which couples to the newly gauged $\tilde{A}_2$ field associated with the $U$ symmetry of the first node. Since this new monopole symmetry is the same on both sides of the duality relating B" and C', when we ultimately arrive at theory C' we have a monopole-like symmetry which still couples to the newly dynamical $\tilde{A}_2$. From theory C' onward, the nodes and links associated with such a symmetry are untouched. Thus the $U(1)$ global symmetry which coupled to the $Y_{1,2}$ bifundamental in theory A becomes a monopole symmetry coupled to the Abelian part of the first $U$ node in theory E.

\begin{figure}
\begin{centering}
\includegraphics[scale=0.6]{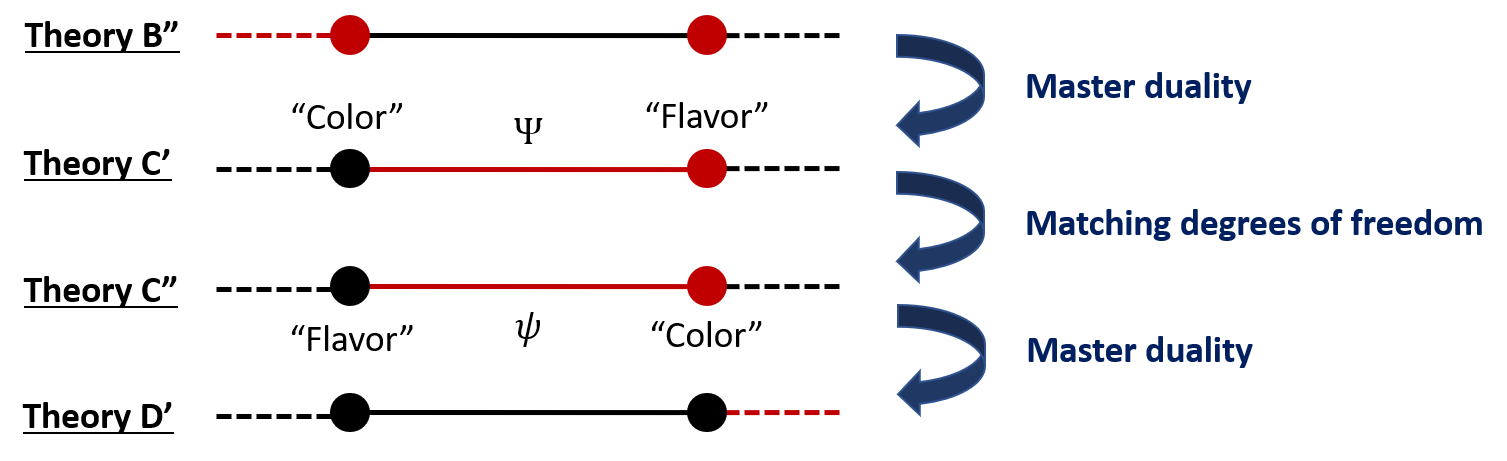}
\par\end{centering}
\caption{How the matching occurs on a generic internal link, which for the purposes of concreteness we have labeled C' and C". \label{fig:matching}}
\end{figure}

Similarly matching must occur for theory C. Fortunately, we don't need to work very hard because the matching between C' and C" is identical to what occurs above for B' and B" and is schematically shown in Fig. \ref{fig:matching}. The additional global symmetry associated with $\tilde{A}_1$ is identified with the global $U(1)$ symmetry on the next node.

We can again follow a global symmetry through from theory A to E for any of the global $U(1)$ symmetries which couple to bifundamentals higher up the quiver. We find results identical to those of the first node/link above: each $U(1)$ global symmetry which couples to the $Y_{i,i+1}$ bifundamental becomes a $U(1)$ global symmetry associated with monopole number for the $i$th node.

Finally, consider how the matching occurs for the last node. In going from theory D" to theory E in the example we considered in the main text, we used \eqref{eq:aharony dual tw}. Since for this duality we need to promote the entire $U(k_2)$ flavor symmetry to dynamical, we once more acquired an additional global symmetry which couples to the newly promoted $\tilde{A}_1$ whose associated background field we call $\tilde{B}_1$. The matching of symmetries is slightly different and is shown in Table \ref{tab:matching_theory_B}. Since Aharony's dualities only have one $U(1)$ global symmetry, there is no monopole-like symmetry associated with the right-most node.

\begin{table}
\begin{centering}
\begin{tabular}{|c|c|c||c|c||c|c|}
\hline
• & B' Side & B" Side & C' Side & C" Side & D' Side & D" Side \\
\hline
$U$ Node & $c$ & $C-\tilde{A}_2$ & $c$ & $C-\tilde{A}_2$ & $c$ & $C+\tilde{A}_1$ \\
$SU$ Node & $B$ & $b^\prime$ & $B$ & $b^\prime$ & $B$ & $b^\prime$ \\
$U(1)_m$ Global Symmetry & $\tilde{A}_1$ & $\tilde{B}_2$ & $\tilde{A}_1$ & $\tilde{B}_2$ & $\tilde{A}_1$ & $\tilde{B}_1$ \\
\hline
\end{tabular}
\par\end{centering}
\caption{Matching of the intermediate theories. Note the matching of theory B' and B" generalizes for any internal matching, except for the very last. Here, $\tilde{B}_2$ is the background gauge field associated with the new global symmetry we get from gauging $\tilde{A}_2$. \label{tab:matching_theory_B}}
\end{table}

We should also point out a special feature of the global symmetries that occurs for certain mass deformed phases. Note that the $\tilde{A}_1$ coupling of \eqref{eq:master_LU} only couples to the unbroken part of the dynamical gauge field $c$. That is, when the $U$ gauge group is broken down to say $U(k_1-k_2)$, the coupling changes from $\text{Tr}_{k_1}(c)dA_1\to\text{Tr}_{k_1-k_2}(c)dA_1$. This is important because when we are in certain mass deformed phases, we must be careful what gauge components are coupled to the $\tilde{A}_1$ charge. Of particular concern in the main text is whether or not vortices couple to certain global symmetries. Since certain finite mass vortices are charged under the broken part of certain gauge fields, they will not couple to particular global symmetries and this will be important for matching excitations.

\subsection{Bosonizing Interaction Terms}
\label{appendix:int}

\begin{figure}
\begin{centering}
\includegraphics[scale=0.6]{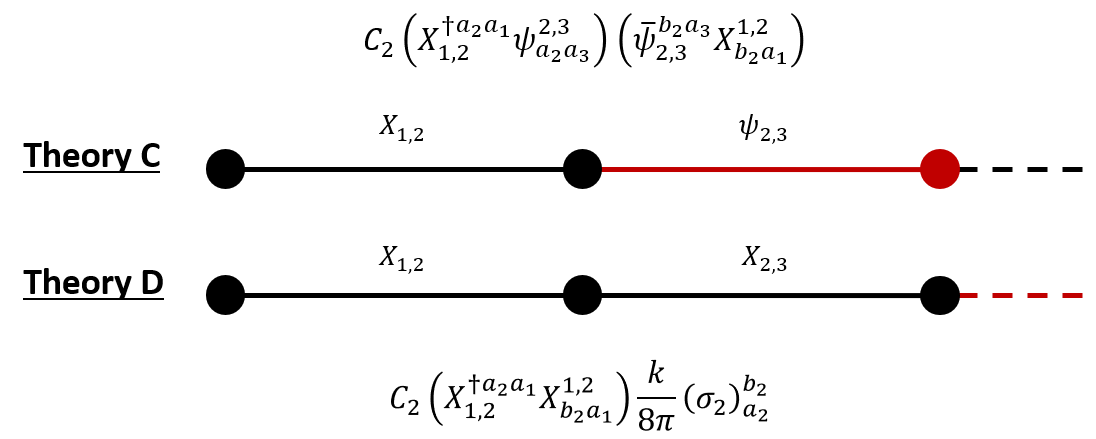}
\par\end{centering}
\caption{Example case of properly dualizing interaction terms. This is for the second node of the four node case considered in the main text in Sec. \ref{sec:building_quivers} and also pictures in Figure \ref{fig:linear quiver tw}. \label{fig:ints}}
\end{figure}

In this appendix, our goal is to justify the proper mapping of the interaction term present in the master duality. We take as an example what occurs on the second node of our four node example in Sec. \ref{sec:building_quivers}, which has a $U(K_2)$ gauge symmetry on it. Specifically, we look at how the interaction term changes when moving from Theory C to Theory D (see Fig. \ref{fig:ints}). With the proper transformation in hand, we will then generalize to interaction terms on the $SU$ and $U$ sides of the duality.

The term from the master duality we would like to apply the duality toward is
\begin{equation}
C_{2}^{(C)}\left(\bar{\psi}_{2,3}^{a_2 a_3}X_{a_2 a_1}^{1,2}\right)\left(X_{1,2}^{\dagger b_2 a_1}\psi_{b_2 a_3}^{2,3}\right).\label{eq:abmm int}
\end{equation}
Although we commit to a magnitude for $C_{2}^{(C)}$ in the main text, to keep this appendix general we only assume $C_{2}^{(C)}<0$ in what follows. From the master duality, the purpose of this interaction term is that when the $X_{1,2}$ field obtains a vacuum expectation value this should cause a subset of the fermions to get a mass \cite{Benini:2017aed,Jensen:2017bjo}.

One might guess the proper transformation of the term between $\psi_{2,3}$ and $X_{2,3}$ from theory C to theory D is simply a generalization of the $m_{\psi} \bar{\psi}\psi\leftrightarrow -m_X^2 \left|X\right|^2$ mass identification. Hence a naive generalization and the shorthand used in the main text is
\begin{equation}
\label{eq:int_naive}
C_{2}^{(C)}\left(\bar{\psi}_{2,3}^{a_2 a_3}X_{a_2 a_1}^{1,2}\right)\left(X_{1,2}^{\dagger b_2 a_1}\psi_{b_2 a_3}^{2,3}\right)
\quad\leftrightarrow\quad
-C_{2}^{(D)}\left(X_{2,3}^{a_2 a_3}X_{a_2 a_1}^{1,2}\right)\left(X_{1,2}^{\dagger b_2 a_1}X_{b_2 a_3}^{2,3}\right).
\end{equation}
This term has the correct behavior when $X_{1,2}$ acquires a vacuum expectation value. Namely, it gives $X_{2,3}$ a color-breaking vacuum expectation value. Unfortunately this suffers from the fact that the reverse procedure can happen as well. That is, when $X_{2,3}$ acquires a vacuum expectation, this gives a mass to $X_{1,2}$. This is because $X_{1,2}$ and $X_{2,3}$ enter on symmetric footing in \eqref{eq:int_naive}, so it is difficult to see where the desired unidirectional behavior that was present in \eqref{eq:abmm int} will come from. This means the TFT deformations do not match, so we must look for a better generalization.

To do so, it will be helpful to go back to the large $N$ and $k$ studies where the duality map between operators is better established. We will assume such mappings continue to hold for finite values of $N$ and $k$, at least up to order one factors, which will not affect the results of our analysis.

Ultimately we'd like to find the true bosonic dual of $\bar{\psi}_{2,3}^{a_2 a_3}\psi_{b_2 a_3}^{2,3}$ with $a_2$ and $b_2$ promoted flavor indices. To do so, we'll look for the bosonic dual of $\bar{\psi}\psi$ and assume our results generalize to arbitrary flavor indices. We follow ref. \cite{Minwalla:2015sca} which represents the most general fermion Lagrangian describing the fixed point as,\footnote{We use a different regularization convention for the Chern-Simons-matter theories than that in \cite{Minwalla:2015sca}. See \cite{Aharony:2015mjs} for a nice discussion on the differences in regularization conventions.}
\begin{align}
\mathcal{L}_{F} & = i\bar{\psi}\Dslash_{b^\prime} \psi-i\left[\frac{N_f-k}{4\pi}\text{Tr}_{N}\left(b^\prime d b^\prime-i\frac{2}{3}b^{\prime 3}\right) \right] \nonumber \\
 & +\sigma_{F}\left(\bar{\psi}\psi-y_2^2\frac{N_f-k}{4\pi}\right)-y_4\frac{N_f-k}{4\pi}\sigma_{F}^{2}+y_6\frac{N_f-k}{4\pi}\sigma_{F}^{3}.
\end{align}
where $\sigma_F$ is an auxiliary field and $y_2^2$, $y_4$, and $y_6$ are arbitrary coefficients of the relevant and marginal operators of the UV Lagrangian. For the coefficients we choose the values
\begin{equation}
y_{4}=x,\qquad y_{2}^{2}=-2x m_\psi,\qquad y_{6}=0.
\end{equation}
Eventually we would like to flow to the deep IR, which amounts to taking $x\to\infty$, where the fermions pick up a finite mass $m_\psi$, but for now we will assume it to be finite. The dual of $\mathcal{L}_F$ on the bosonic side is
\begin{align}
\mathcal{L}_{B} & =|D_{c}\phi|^2-i\left[\frac{N}{4\pi}\text{Tr}_k\left(cdc-i\frac{2}{3}c^3\right)\right] \nonumber \\
 & +m_{B}^{2}\phi^{\dagger}\phi+b_4\frac{4\pi}{N}\left(\phi^{\dagger}\phi\right)^{2}+x_6\frac{\left(2\pi\right)^{2}}{N^2}\left(\phi^{\dagger}\phi\right)^{3}
\end{align}
where $m_B^2$, $b_4$, and $x_6$ are coefficient of marginal/relevant operators (the dual of the fermionic parameters) which are then
\begin{equation}
b_{4}=x,\qquad m_{B}^{2}=2x\left(\frac{N-k}{N}\right) m_\phi^2,\qquad x_{6}=0.
\end{equation}

Unfortunately, in the analysis of ref. \cite{Minwalla:2015sca}, there is no clear dual to $\bar{\psi}\psi$. We can however express $\bar{\psi}\psi$ in terms of operators which do have a well-defined dual. Integrating out $\sigma_F$ in $S_F$ enforces the equations of motion
\begin{equation}
x\frac{N_f-k}{2\pi}\sigma_{F}=\bar{\psi}\psi+x\frac{N_f-k}{2\pi}m_\psi.
\end{equation}
We can rearrange this expression to solve for $\bar{\psi}\psi$,
\begin{equation}
\bar{\psi}\psi=x\frac{N_f-k}{2\pi}\left(\sigma_{F}-m_\psi\right).
\end{equation}
Fortunately, we know how to dualize the RHS from the maps in \cite{Minwalla:2015sca}. We obtain
\begin{equation}
\bar{\psi}\psi=2x\frac{N_f-k}{4\pi}\left(\sigma_{F}-m_\psi\right)\leftrightarrow-2x\left(\phi^{\dagger}\phi+\frac{N-k}{4\pi}m_\phi^2\right).\label{eq:abmm psibpsi}
\end{equation}

Now let's see how this is implemented for the WF scalar. We know we can rewrite the WF scalar using auxiliary fields, namely, we write the quadratic and quartic terms so that $S_B$ then contains the terms
\begin{equation}
S_{B}\supset\int d^{3}x\,\left[\sigma_{B}\left(\phi^{\dagger}\phi+\frac{N-k}{4\pi}m_\phi^2\right)-\frac{N}{16\pi x}\sigma_{B}^{2}\right].\label{eq:abmm Sb}
\end{equation}
Again, $\sigma_B$ is an auxiliary field and its corresponding equation of motion is
\begin{equation}
\phi^{\dagger}\phi+\frac{N-k}{4\pi}m_\phi^2=\frac{N}{16\pi x}\sigma_{B}.
\end{equation}
Notably, the LHS of this expression matches the term on the RHS of \eqref{eq:abmm psibpsi}, so we can establish the duality
\begin{equation}
\bar{\psi}\psi\leftrightarrow-\frac{N}{8\pi}\sigma_{B}.\label{eq:abmm dual}
\end{equation}
Note all factors of $x$ have dropped from this expression.

Hence, under the bosonization of the fermion end of this interaction, a naive generalization of \eqref{eq:abmm dual} for \eqref{eq:abmm int} would be
\begin{equation}
C_2^{(C)}\left(\bar{\psi}_{2,3}^{a_2 a_3}X_{a_2 a_1}^{1,2}\right)\left(X_{1,2}^{\dagger b_2 a_1}\psi_{b_2 a_3}^{2,3}\right)
\quad\leftrightarrow\quad
-C_2^{(D)}\left(X_{1,2}^{\dagger b_2 a_1}X_{a_2 a_1}^{1,2}\right) \frac{N}{8\pi} \left(\sigma_{2,3}\right)_{b_2}^{a_2}
\end{equation}
where we have introduced the $K_2\times K_2$ auxiliary fields $\sigma_{2,3}$. Importantly, there is an inherent asymmetry here because $\sigma_{2,3}$ belongs to the $X_{2,3}$ link. Grouping this together with the other terms linear in $\sigma_{2,3}$ and taking the $x\to\infty$ limit, the bosonic action of the link to the right of the node is a generalization of \eqref{eq:abmm Sb} is
\begin{equation}
S_{B}\supset\int d^{3}x\,\left(\sigma_{2,3}\right)_{b_2}^{a_2}\left(X_{2,3}^{\dagger b_2 a_3}X^{2,3}_{a_2 a_3}+\frac{N-k}{4\pi}m^2_{2,3}\delta_{a_2}^{b_2}-C^{(D)}_{2}\frac{N}{8\pi}X_{1,2}^{\dagger b_2 a_1}X_{a_2 a_1}^{1,2}\right)\label{eq:abmm new int}
\end{equation}
where we have assumed that we have used the $U(K_2)$ symmetry such that $\left(m^2\right)_{a_2}^{b_2}$ is always diagonal.

Before proceeding, let us briefly comment on the need to introduce $(K_2)^2$ $\sigma$ fields. Had we introduced only $K_2$ sigma fields, all self-interactions would have been of the form 
\begin{equation} 
\sum_{a_2} \left(X_{2,3}^{\dagger a_2 a_3}X_{a_2 a_3}^{2,3}\right)^{2}.
\label{eq:flavorbroken}
\end{equation}
In contrast, \eqref{eq:abmm new int}
implies self-interaction terms of the form $\left(X_{2,3}^{\dagger b_2 a_3}X_{a_2 a_3}^{2,3}\right)^{2}$
with \emph{no sum} over $a_2$ and $b_2$. These can be combined into a perfect square, $\left( \sum_{a_2} X_{2,3}^{\dagger a_2 a_3}X_{a_2 a_3}^{2,3}\right)^{2}$. This potential is needed to realize the full ``flavor'' symmetry, i.e. $SU(N_s)$. Meanwhile, \eqref{eq:flavorbroken} only preserves the diagonal $U(1)^{N_s}$ subgroup. For example, in the $N_s=2$ the two distinct choices for the interaction term are
\begin{align}
\left|\phi_{1}\right|^{4}+\left|\phi_{2}\right|^{4}\qquad\text{or}\qquad\left(\left|\phi_{1}\right|^{2}+\left|\phi_{2}\right|^{2}\right)^{2}.
\end{align}
The former is invariant under a $U(1)^{2}$ symmetry while the latter is invariant under $SU(2)$. While it doesn't appear that anyone has been careful enough to distinguish the two possibilities in the context of Non-Abelian 3d bosonization dualities\footnote{In the Abelian case, dualities which differ only by the structure of the quartic interactions have been discussed in \cite{Wang:2017txt}.}, it appears that for symmetry matching of the global flavor symmetries we do need \eqref{eq:abmm new int} with its $(K_2)^2$ auxiliary $\sigma$ fields.

To analyze the effect of the interaction terms in \eqref{eq:abmm new int} , let us recall how the original WF term without the additional interactions, $\mathcal{L}_{B}\supset\sigma_{B}\left(\phi^{\dagger} \phi+\frac{N-k}{4\pi}m_\phi^2\right)$. We know that we should expect
\begin{subequations}
\label{eq:mphi_vevs}
\begin{align}
m_\phi^2>0:\qquad & \left\langle \phi^{\dagger} \phi\right\rangle =0\\
m_\phi^2<0:\qquad & \left\langle \phi^{\dagger} \phi\right\rangle \sim m_\phi^2.
\end{align}
\end{subequations}
We can use this to make conclusions about \eqref{eq:abmm new int}. Note that we are assuming the $U$ groups on the nodes are of unequal rank and thus $\left\langle X^{\dagger a_1 a_2}_{1,2} X^{1,2}_{a_1 b_2}\right\rangle = (v^2_{1,2})^{a_2}_{b_2}\ne v^2\delta^{a_2}_{b_2}$ as in \eqref{eq:u vev tw} for some constant $v$. We'll also assume maximal Higgsing so that the a negative deformation of $m^2_{2,3}$ produces a vacuum expectation value of $\left\langle X^{\dagger a_2 a_3}_{2,3} X^{2,3}_{b_2 a_3}\right\rangle = (v^2_{2,3})^{a_2}_{b_2} \sim \delta^{a_2}_{b_2}$. We then find
\begin{subequations}
\begin{align}
m_{2,3}^2>0,\qquad\left\langle X^{\dagger a_1 a_2}_{1,2} X^{1,2}_{a_1 b_2}\right\rangle >0:\qquad & \left\langle X^{\dagger a_2 a_3}_{2,3} X^{2,3}_{b_2 a_3}\right\rangle \sim C_2^{(D)}\left(v_{1,2}\right)^{a_2}_{b_2}\\
m_{2,3}^2>0,\qquad\left\langle X^{\dagger a_1 a_2}_{1,2} X^{1,2}_{a_1 b_2}\right\rangle =0:\qquad & \left\langle X^{\dagger a_2 a_3}_{2,3} X^{2,3}_{b_2 a_3}\right\rangle =0\\
m_{2,3}^2<0,\qquad\left\langle X^{\dagger a_1 a_2}_{1,2} X^{1,2}_{a_1 b_2}\right\rangle >0:\qquad & \left\langle X^{\dagger a_2 a_3}_{2,3} X^{2,3}_{b_2 a_3}\right\rangle \sim \left(C_2^{(D)} v_{1,2}+v_{2,3}\right)^{a_2}_{b_2}.\\
m_{2,3}^2<0,\qquad\left\langle X^{\dagger a_1 a_2}_{1,2} X^{1,2}_{a_1 b_2}\right\rangle =0:\qquad & \left\langle X^{\dagger a_2 a_3}_{2,3} X^{2,3}_{b_2 a_3}\right\rangle \sim \left(v_{2,3}\right)^{a_2}_{b_2}.
\end{align}
\end{subequations}
This is precisely the behavior that we would expect for a term dual to \eqref{eq:abmm int}. That is, we see that a nonzero vacuum expectation value of the $X_{1,2}$ fields can cause certain components of $X_{2,3}$ to also get a vacuum expectation value. Importantly, if $X_{2,3}$ has a vacuum expectation value but $X_{1,2}$ does not, the interaction term does not work in reverse. To see this, note that the equation of motion for $(\sigma_2)^{a_2}_{b_2}$ would require
\begin{equation}
\label{eq:x23_eom}
(v^2_{2,3})^{a_2}_{b_2}+(N-k)\frac{m_{2,3}^2}{4\pi}=C^{(D)}_{2}\frac{N}{8\pi}X_{1,2}^{\dagger b_2 a_1}X_{a_2 a_1}^{1,2}.
\end{equation}
Similar equations of motion hold for the $X_{3,4}$ link, but importantly the $X_{1,2}$ link contains no such interaction term. To determine the vacuum expectation values, one then just needs to consistently solve the three equations of motion. To do so, it is helpful to start with the one corresponding to the $X_{1,2}$ link. Since there is no interaction term for the $X_{1,2}$ link, $X_{2,3}$ has no influence on $v_{1,2}$ and its value follows in a manner completely analogous to \eqref{eq:mphi_vevs}. When $X_{1,2}$ acquires a vacuum expectation value, it then affects the $X_{2,3}$ equation of motion via \eqref{eq:x23_eom}. Hence we get our desired unidirectional influence. This behavior was always difficult to achieve using the naive interaction generalization in \eqref{eq:int_naive} because both the $X_{2,3}$ and $X_{1,2}$ fields appeared in the term symmetrically. Thus the effect of one field acquiring a nonzero vacuum expectation value was always the same as the other field, up to flavor structure.

More generally, for interactions on the $U$ side we have
\begin{equation}
\label{eq:u_ints}
C_I\left(\bar{\psi}_{I,I+1}^{a_I a_{I+1}}X_{a_I a_{I-1}}^{I-1,I}\right)\left(X_{I-1,I}^{\dagger b_I a_{I-1}}\psi_{b_I a_{I+1}}^{I,I+1}\right)
\quad\leftrightarrow\quad
-C_I\left(X_{I-1,I}^{\dagger b_I a_{I-1}}X_{a_I a_{I-1}}^{I-1,I}\right)\frac{K_{I-1}}{8\pi} \left(\sigma_{I,I+1}\right)_{b_I}^{a_I}
\end{equation}
while for interactions on the $SU$ side, which are unidirectional to the left,
\begin{equation}
\label{eq:su_ints}
C_I \left(\bar{\psi}_{I-1,I}^{a_{I-1} a_I}Y_{a_I a_{I+1}}^{I,I+1}\right)\left(Y_{I,I+1}^{\dagger b_{I+1} a_I}\psi_{b_{I+1} a_I}^{I-1,I}\right)
\quad\leftrightarrow\quad
-C_I \left(Y_{I,I+1}^{\dagger b_{I+1} a_I}Y_{a_{I+1} a_I}^{I,I+1}\right)\frac{N_I}{8\pi} \left(\sigma_{I-1,I}\right)_{b_I}^{a_I}
\end{equation}
where the $\sigma_{I,I+1}$ are auxiliary fields belonging to the $(I,I+1)$-th bifundamental and $I$ runs over all internal nodes (e.g. $I=2,3$ for the four node case).

\bibliographystyle{JHEP}
\bibliography{domainwallstrings}
\end{document}